\documentclass[pdflatex,sn-nature]{sn-jnl}

\UseRawInputEncoding
\usepackage{graphicx}
\usepackage{multirow}
\usepackage{amsmath,amssymb,amsfonts}
\usepackage{amsthm}
\usepackage{mathrsfs}
\usepackage[title]{appendix}
\usepackage{xcolor}
\usepackage{textcomp}
\usepackage{manyfoot}
\usepackage{booktabs}
\usepackage{listings}
\usepackage{subcaption}
\usepackage{float}
\usepackage{hyperref}
\usepackage{longtable}
\usepackage{pdflscape}
\usepackage{eso-pic}
\usepackage{array}
\usepackage{tabularx}

\newcolumntype{Y}{>{\raggedright\arraybackslash}X}
\newcolumntype{C}[1]{>{\centering\arraybackslash}p{#1}}

\theoremstyle{thmstyleone}

\theoremstyle{thmstyletwo}

\theoremstyle{thmstylethree}

\begin{document}
	
	\title[Frequency-consistent synthesis and selective transfer]
	{Frequency-consistent synthesis and selective transfer for medical image analysis}
	
	\author[1,2]{\fnm{Zhuhui} \sur{Wang}}
	\email{wangzhuhui@stu.pku.edu.cn}
	
	\author[1,2]{\fnm{Jianwei} \sur{Zuo}}
	\email{zuojianwei@stu.pku.edu.cn}
	
	\author[3,4]{\fnm{Xin} \sur{Peng}}
	\email{pxpengxin@263.net}
	
	\author*[1,2]{\fnm{Kai} \sur{Zhong}}
	\email{kzhong@bjmu.edu.cn}
	
	\author*[2,5,6]{\fnm{Xuliang} \sur{Deng}}
	\email{kqdengxuliang@bjmu.edu.cn}
	
	\author*[1,2]{\fnm{Jiajia} \sur{Luo}}
	\email{jiajia.luo@pku.edu.cn}
	
	\affil[1]{
		\orgdiv{Institute of Medical Technology},
		\orgname{Peking University Health Science Center},
		\orgaddress{
			\city{Beijing},
			\postcode{100191},
			\country{China}}}
	
	\affil[2]{
		\orgdiv{Biomedical Engineering Department, Institute of Advanced Clinical Medicine},
		\orgname{Peking University},
		\orgaddress{
			\city{Beijing},
			\postcode{100191},
			\country{China}}}
	
	\affil[3]{
		\orgdiv{Department of Oral and Maxillofacial Surgery},
		\orgname{Peking University School and Hospital of Stomatology},
		\orgaddress{
			\city{Beijing},
			\postcode{100081},
			\country{China}}}
	
	\affil[4]{
		\orgdiv{National Clinical Research Center for Oral Diseases, National Engineering Laboratory for Digital and Material Technology of Stomatology},
		\orgname{Beijing Key Laboratory of Digital Stomatology},
		\orgaddress{
			\city{Beijing},
			\postcode{100081},
			\country{China}}}
	
	\affil[5]{
		\orgdiv{National Engineering Research Center of Oral Biomaterials and Digital Medical Devices, NMPA Key Laboratory for Dental Materials, Beijing Laboratory of Biomedical Materials},
		\orgname{Peking University School and Hospital of Stomatology},
		\orgaddress{
			\city{Beijing},
			\postcode{100081},
			\country{China}}}
	
	\affil[6]{
		\orgdiv{Department of Geriatric Dentistry},
		\orgname{Peking University School and Hospital of Stomatology},
		\orgaddress{
			\city{Beijing},
			\postcode{100081},
			\country{China}}}
	
	\abstract{
		Complementary imaging is not always available, and realistic synthesis does not necessarily improve downstream analysis. Here we introduce a training framework linking medical image synthesis to task learning through adaptive frequency consistency and selective feature transfer. Image-specific frequency constraints encourage consistency of structure and detail. Features are transferred only when a teacher using synthetic images achieves a lower task-error score than the student by a task-specific margin on a labelled training sample. Evaluations across five datasets show improved translation fidelity with three adversarial and diffusion-based translators trained without paired supervision, alongside gains in brain-tumour segmentation on tumour-containing slices and lesion centre-point detection. All 105 reported paired comparisons yielded point estimates favouring the proposed methods, including 36 external translation comparisons on post-treatment MRI without retraining. These findings support synthetic modalities as task-oriented complements to acquired scans when complementary imaging is unavailable.
	}
	
	\maketitle
	
	\section*{Main}\label{sec1}
	
	\par Complementary imaging modalities and sequences provide different views of anatomy and pathology, supporting diagnosis, treatment planning and monitoring \cite{esr2015personalised,james2014fusion}. Computed tomography (CT) and magnetic resonance imaging (MRI), for example, offer distinct forms of tissue contrast \cite{radiologyinfo_bodyct,1,nibib_mri,2}, but acquiring all relevant images can be impractical because of examination time and cost \cite{radiologyinfo_chestmri,nibib_mri}. Image synthesis offers a computational means of deriving an additional representation from an acquired scan without another acquisition. For downstream analysis, however, resemblance to a target modality is not enough: the synthesized representation must also support the task for which it is used \cite{5,6}. This motivates an approach that addresses both image fidelity and how synthetic representations enter task learning.
	
	\par Medical image translation methods based on generative adversarial networks (GANs) and diffusion models enable synthesis across modalities \cite{gan,ddpm,ddim,ldm,syndiff,cvpr18,3}. Frequency-aware objectives provide additional constraints, but fixed decompositions do not adapt frequency separation to the characteristics of individual images \cite{fdit,li2023zero,gao2024frequency}. A separate challenge concerns how synthetic representations are used. Knowledge distillation transfers learned representations \cite{hinton2015distilling,romero2015fitnets}, and image enhancement has also been linked to improved downstream analysis \cite{sun2025foundation}. Yet neither image fidelity nor feature transfer alone establishes that a teacher provides a task-specific advantage for every training sample. Building on joint translation--segmentation methods \cite{cvpr18} and representation transfer \cite{hinton2015distilling,romero2015fitnets,li2024multistage,pan2009survey}, we therefore investigate a framework combining image-adaptive frequency consistency with feature transfer conditioned on the relative task performance of a teacher using synthetic images.
	
	\par Here we introduce a training framework that links these two forms of adaptation. Frequency-Domain Decomposition Translation (FDDT) encourages consistency between translations of low- and high-frequency inputs and the corresponding components of the full-image translation \cite{wang2023fddt,fdit,li2023zero,gao2024frequency}. Adaptive Frequency Decomposition (AFD) learns image-specific cutoffs using a spatial--frequency cross-attention module rather than imposing a single fixed partition \cite{ca-attn,vaswani2017attention}. Selective Feature Distillation (SFD) then uses a teacher operating on synthetic images to guide a student receiving both acquired and synthetic images. Feature transfer is activated only when the teacher has a lower task-error score than the student by a task-specific margin on a labelled training sample. The framework extends our earlier GAN-based FDDT formulation \cite{wang2023fddt} with adaptive decomposition, selective downstream learning and evaluation across adversarial and diffusion-based translators.
	
	\begin{figure}[H]
		\centering
		\includegraphics[width=0.90\textwidth]{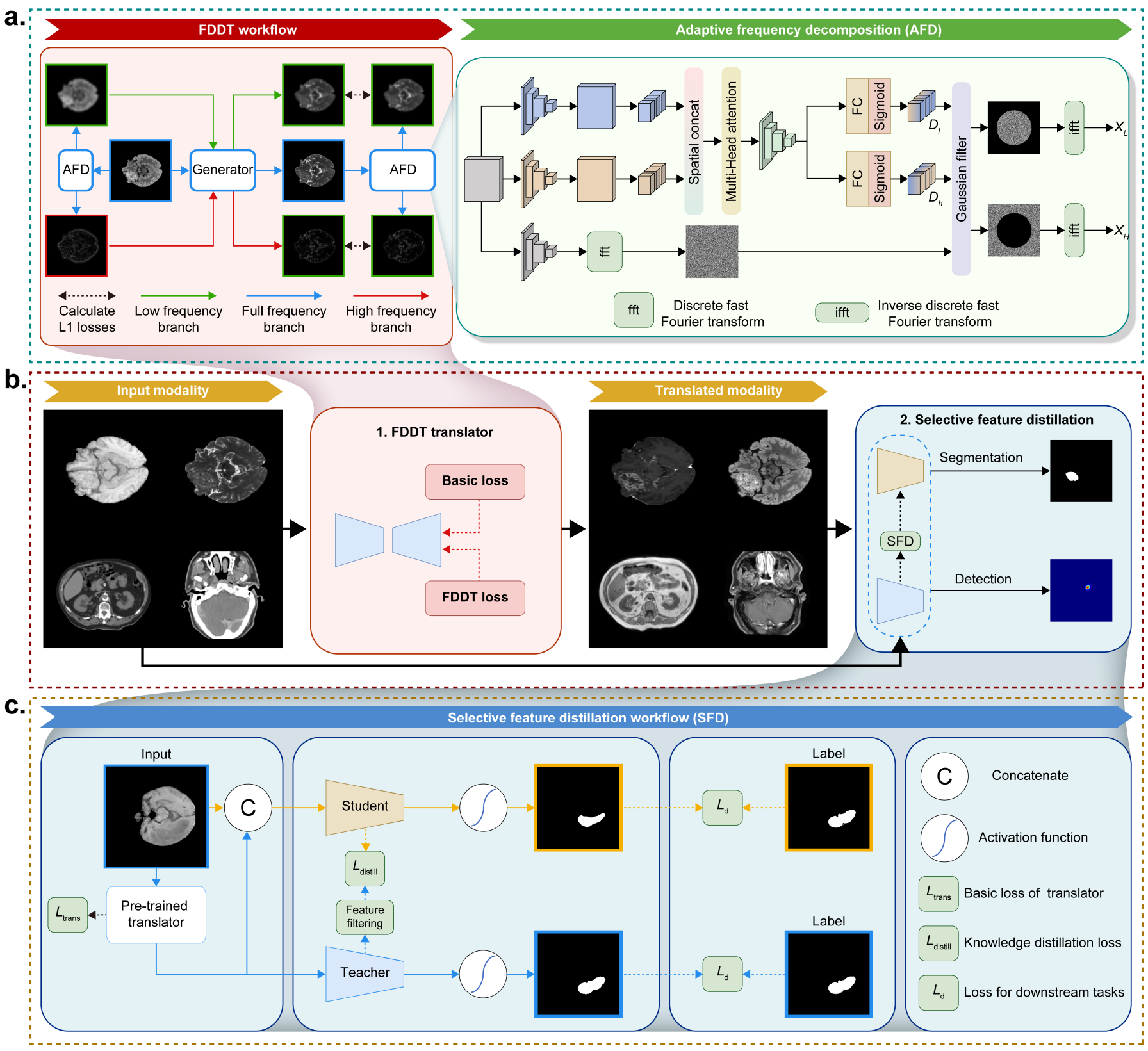}
		\caption{
			\textbf{Frequency-consistent image translation and selective feature distillation.}
			\textbf{(a)} During FDDT training, the shared base translator processes full-, low-, and high-frequency inputs. A single shared AFD module predicts sample-specific cutoffs for the source and generated images, and frequency-consistency losses align translations of the decomposed inputs with the corresponding components of the full-image translation. The frequency pathways and AFD module are removed after training.
			\textbf{(b)} The generated modality is combined with the source modality for downstream segmentation and lesion centre-point detection.
			\textbf{(c)} During SFD training, feature distillation is activated only when the teacher has a lower task-specific risk than the student by the predefined margin, while a low-weight translation GAN objective remains active to regularize the jointly fine-tuned translator. At inference, the translator and student are retained, while the teacher and distillation branches are removed.
		}
		\label{fig1}
	\end{figure}
	
	\par We evaluate the framework across five imaging datasets through complementary experiments on translation fidelity, adaptive decomposition, downstream utility and external translation performance. Translation is tested with CycleGAN, CUT and SynDiff under unpaired training on BraTS 2021 and SynthRAD2025 \cite{cyclegan,cut,syndiff,brats2021_1,synthrad2025}, and adaptive decomposition is examined using bidirectional PD--T2 translation on IXI \cite{ixi_dataset}. Downstream evaluations comprise whole-tumour segmentation on tumour-containing slices from BraTS 2021 and lesion centre-point detection on DeepLesion \cite{unet,centernet,deeplesion}. Separately, translators trained on preoperative MRI are evaluated without retraining on longitudinal post-treatment MRI from BraTS 2024 to assess whether translation gains persist under distribution shift \cite{brats2024_2}. All test sets are held out at the case or patient level.
	
	\par This design treats synthetic modalities as task-oriented complements to acquired scans rather than assuming that synthesis alone establishes their downstream value. Selection governs teacher-feature transfer during supervised training, not whether a synthetic image is accepted at inference; the student continues to receive both acquired and synthetic inputs. The auxiliary frequency, AFD, teacher and distillation components operate only during training. At inference, the downstream pipeline retains the translator and student, but not the teacher or distillation branches.

	\section*{Results}\label{sec:results}
	
	\subsection*{Frequency-consistent synthesis}
	
	\par FDDT augments a shared translator with full-, low- and high-frequency training pathways and enforces consistency between frequency-band translations and the corresponding components of the full-image translation (Fig.~\ref{fig1}) \cite{wang2023fddt,fdit,li2023zero,gao2024frequency}. AFD adapts the frequency cutoffs to each image using spatial and spectral representations \cite{ca-attn,vaswani2017attention}. For downstream learning, SFD conditions teacher-feature transfer on the teacher--student task-error comparison. The auxiliary frequency and distillation branches are removed after training; the deployed downstream pipeline retains the translator and the student receiving acquired and synthetic images. Unless otherwise specified, FDDT in Figs.~\ref{fig2}, \ref{fig4}, and \ref{fig5} refers to FDDT incorporating AFD during translator pretraining, whereas Fig.~\ref{fig3} explicitly distinguishes fixed-threshold FDDT from FDDT with AFD.
	
	\par We evaluated FDDT with CycleGAN, CUT and SynDiff, spanning adversarial and diffusion-based unpaired translation \cite{cyclegan,cut,syndiff}. BraTS 2021 models were trained on 1,100 cases and selected using a separate 151-case internal validation partition before final evaluation on the 219-case official validation cohort. Across T1\(\rightarrow\)T1ce and T2\(\rightarrow\)FLAIR translation, FDDT increased mean SSIM and PSNR and reduced mean LPIPS relative to both the unmodified backbones and their FDIT variants (Fig.~\ref{fig2}a) \cite{wang2004image,huynh2008scope,lpips,fdit}. Relative to the unmodified backbones, mean case-level improvements ranged from 0.0167 to 0.0320 for SSIM and from 1.40 to 3.04~dB for PSNR, while LPIPS decreased by 0.0031--0.0167 (Supplementary Table~\ref{tab:stat_fig2a}). Qualitative examples suggested clearer lesion boundaries and anatomical structures. On SynthRAD2025, paired CT--MR data were used for unpaired CT\(\rightarrow\)MR translation, with correspondence withheld from the training objective and retained for partitioning and evaluation. FDDT improved SSIM and PSNR and reduced FID across all three backbones relative to the base models and FDIT variants (Fig.~\ref{fig2}b) \cite{synthrad2025,heusel2017gans}. Relative to the unmodified backbones, mean case-level SSIM increased by 0.0122--0.0175 and PSNR by 0.42--0.57~dB (Supplementary Table~\ref{tab:stat_fig2b}). These results support frequency consistency across the evaluated model families, anatomical regions and modality pairs.
	
	\par Case-level analyses supported the translation results in the 219-case BraTS 2021 and 103-case SynthRAD2025 evaluation cohorts. For the reported SSIM, PSNR and LPIPS comparisons, pointwise 95\% CIs for the mean paired differences excluded zero in the direction favouring FDDT over both the base model and FDIT variant (Supplementary Tables~\ref{tab:stat_fig2a} and~\ref{tab:stat_fig2b}). All comparisons remained significant after Holm correction within their respective table-wise families (Holm-adjusted \(p<0.001\)). FID was calculated at the dataset level and reported descriptively, without paired case-level testing.
	
	\subsection*{Adaptive frequency decomposition}
	
	\par To assess the contribution of image-adaptive frequency separation, we compared the base translator, fixed-threshold FDDT and FDDT with AFD on bidirectional PD--T2 translation in IXI (Fig.~\ref{fig3}b) \cite{ixi_dataset}. AFD uses a shared spatial--frequency predictor to estimate image-specific low- and high-frequency cutoffs within predefined ranges \cite{ca-attn,vaswani2017attention}. Band-specific error maps from BraTS 2021 illustrate correspondence between translated and target frequency components (Fig.~\ref{fig3}a), while the IXI cutoff distributions show the behaviour of the adaptive predictor (Fig.~\ref{fig3}c). The fixed-versus-adaptive comparison provides the direct test of the contribution of cutoff adaptation.
	
	\par Case-level analysis of the IXI test cohort quantified the benefit of adaptive cutoff regression (\(n=88\) paired cases). Paired Wilcoxon signed-rank tests were used. Mean paired differences are reported to align the effect estimates with the cohort-mean bars in Fig.~\ref{fig3}b, while 95\% percentile-bootstrap CIs describe the median paired differences (5,000 resamples; seed 42). Relative to fixed-threshold FDDT, adding AFD improved both SSIM and PSNR for PD\(\rightarrow\)T2 and T2\(\rightarrow\)PD translation (all raw \(p<0.001\); Holm-adjusted \(p<0.01\)). The mean improvements were 0.0137 for SSIM and 0.719~dB for PSNR in PD\(\rightarrow\)T2 translation, and 0.0116 and 1.115~dB, respectively, in T2\(\rightarrow\)PD translation. The corresponding 95\% bootstrap CIs for the median differences were 0.0155--0.0223 and 0.355--1.004~dB for PD\(\rightarrow\)T2, and 0.0098--0.0247 and 0.508--1.214~dB for T2\(\rightarrow\)PD (Supplementary Table~\ref{tab:afd_sig_appendix}). FID was reported descriptively at the dataset level and was not included in the case-level Wilcoxon analysis.
	
	\subsection*{Selective feature transfer}
	
	\par On the 249-case held-out BraTS 2021 segmentation test partition,
	SFD achieved the highest precision, recall, Dice and IoU among
	the evaluated methods for tumour-containing slices (Fig.~\ref{fig4}a).
	Its generator was pretrained independently using only the 900 training
	cases; the 102 validation and 249 test cases were excluded
	from pretraining and downstream optimization. Compared with UNet-SE-CA
	and Zhang et al. \cite{unet,senet,ca-attn,cvpr18}, SFD increased
	mean Dice by 0.0331 and 0.0261, respectively,
	and mean IoU by 0.0417 and 0.0325, respectively. The corresponding pointwise 95\% CIs excluded zero, and all eight segmentation comparisons remained significant after table-wise Holm correction (Holm-adjusted \(p<0.05\); Supplementary Table~\ref{tab:sfd_seg_sig_appendix}).
	
	The ablation experiments in Fig.~\ref{fig4}b,c used a separately defined
	80/16/16 training/validation/test split. For these ablations, the generator
	was pretrained independently using only the corresponding 80-case training
	partition; the 16 validation and 16 test cases were excluded from generator
	pretraining and model optimization. These experiments evaluated loss-gated SFD against ungated full feature distillation and examined the contributions of the base translator, FDIT and FDDT to downstream segmentation performance. These ablations assess the contribution of selective transfer in the separately trained 16-case test setting, rather than constituting an additional test on the 249-case main cohort.
	
	\par On DeepLesion, adding SFD to the CenterNet-style heatmap detector increased patient-wise sensitivity at FP budgets of 0.5, 1, and 2 false positives per image and reduced KL divergence and conditional mean localization error (Fig.~\ref{fig4}e) \cite{deeplesion,centernet}. Relative to CenterNet, SFD increased patient-aggregated sensitivity by 0.0282, 0.0289 and 0.0414 at 0.5, 1 and 2 false positives per image, respectively, and reduced KL divergence by 0.4392 (Supplementary Table~\ref{tab:sfd_det_sig_appendix}). Patient-level analyses included 663 test patients for sensitivity and KL divergence. Among the 528 patients with valid matched detections for both methods, SFD reduced the conditional mean localization error by 0.216 pixels. All five detection comparisons remained significant after Holm correction within family S6 (Holm-adjusted \(p<0.05\)).
	
	\subsection*{External validation}
	
	\par The same locked BraTS 2021 translator checkpoints, optimized on the 1,100-case training partition and selected using the 151-case internal validation partition, were evaluated without retraining or further model selection on 87 patients from the BraTS 2024 validation data, comprising 187 longitudinal post-treatment MRI examinations \cite{brats2024_2,cyclegan,cut,syndiff}. Examination-level SSIM, PSNR and LPIPS values were averaged within patient before analysis. Across the two translation tasks, three translator backbones and two reference methods, all 36 patient-level mean differences favoured FDDT (Fig.~\ref{fig5}a,b). Relative to the unmodified backbones, FDDT increased mean SSIM by 0.0045--0.0101 and PSNR by 0.32--0.98~dB, while reducing LPIPS by 0.0055--0.0102. Overall, 35 of the 36 corresponding pointwise 95\% CIs excluded zero in the direction favouring FDDT; 30 of the 36 comparisons remained significant after Holm correction (Holm-adjusted \(p<0.05\)). For T2\(\rightarrow\)FLAIR, the CUT+FDDT versus CUT+FDIT LPIPS comparison was not significant after correction (Holm-adjusted \(p=0.15595\)). Complete results are reported in Supplementary Table~\ref{tab:brats2024_sig_appendix}. This external evaluation assessed image translation, not downstream segmentation or detection.
	
	\subsection*{Training and inference costs}
	
	\par The computational overhead of the framework differs between its translation and downstream-task components. AFD is implemented as a single module with parameters independent of generator \(G\) and is shared between decomposition of the real input and generated output. This shared module adds 0.27 million trainable parameters and is used only during translation training. With CycleGAN, batch size 1 and an input size of \(1\times256\times256\), the additional full-, low- and high-frequency processing together with AFD increases training computation by approximately 230.94~G MACs per iteration relative to the base training graph. Because the frequency-specific branches and AFD module are removed after optimization, the deployed FDDT translator has the same inference architecture, parameter count and nominal MACs as the corresponding base translator. For SFD, the symmetric teacher and feature-alignment branches are active only during training and approximately double task-network training computation. At inference, translator \(G\) generates \(\tilde{Y}\), and the retained student processes \([X,\tilde{Y}]\). SFD therefore introduces no teacher- or distillation-specific inference overhead relative to the same translator-conditioned student pipeline, although it retains the cost of the translator compared with an \(X\)-only task network.
	
	\begin{figure}[H]
		\centering
		\includegraphics[width=0.95\textwidth]{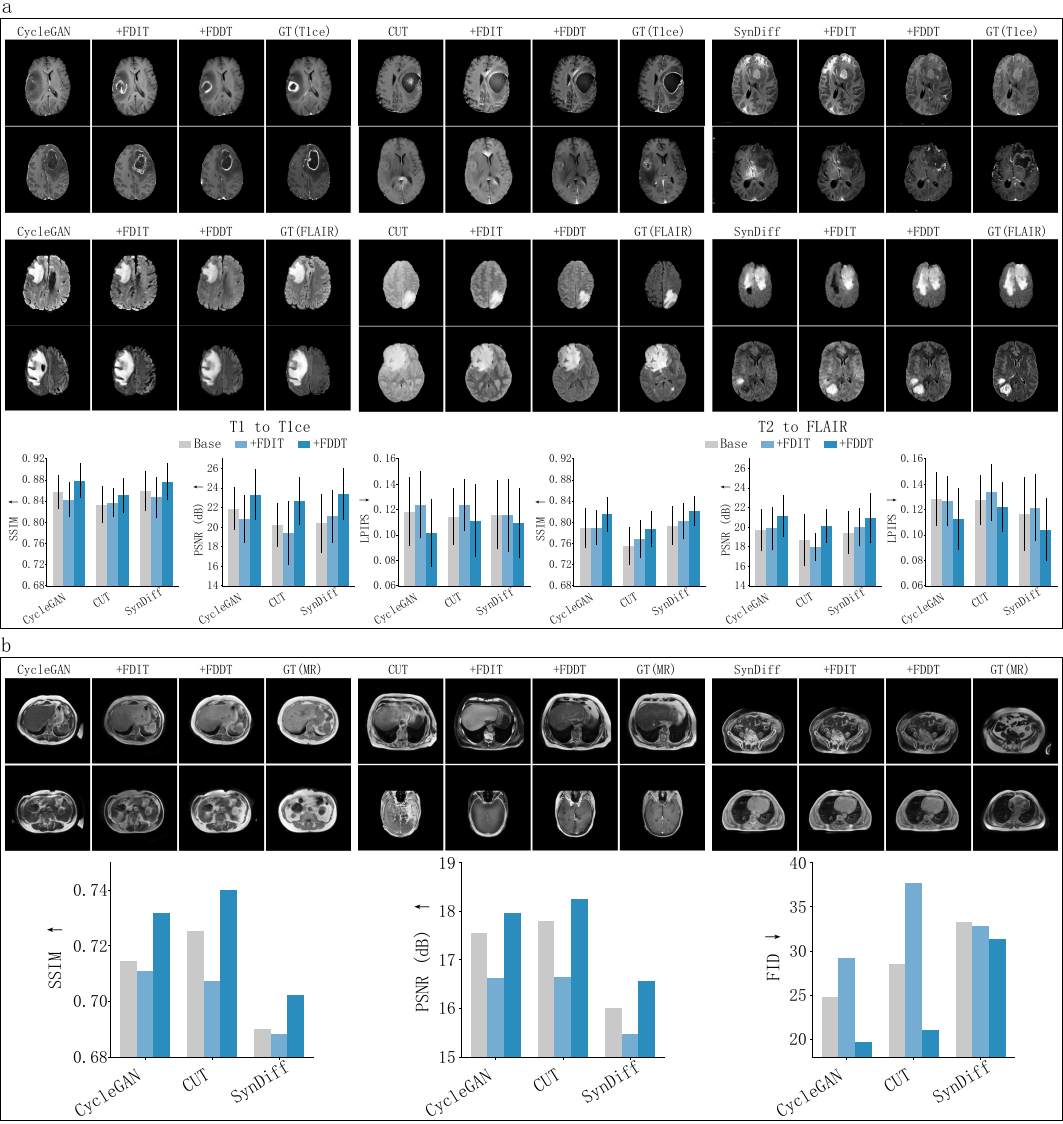}
		\caption{
			\textbf{Cross-modal image-translation performance.}
			\textbf{(a)} Representative T1\(\rightarrow\)T1ce and T2\(\rightarrow\)FLAIR translations and case-level SSIM, PSNR and LPIPS results for the 219-case BraTS 2021 official validation cohort, used here exclusively for final evaluation.
			\textbf{(b)} Representative CT\(\rightarrow\)MR translations, case-level SSIM and PSNR results, and dataset-level FID for the independently held-out 103-case SynthRAD2025 test partition. GT denotes the paired target-modality image, used to evaluate the corresponding held-out source image; source--target correspondence was not provided to the training objective. For the case-level metrics in \textbf{(a)}, bars show cohort means and error bars indicate standard deviations (SDs) across cases. In \textbf{(b)}, SSIM and PSNR bars show case-level cohort means without error bars, while FID is reported as a dataset-level value. Raw and Holm-adjusted results are reported in Supplementary Tables~\ref{tab:stat_fig2a} and~\ref{tab:stat_fig2b}.
		}
		\label{fig2}
	\end{figure}
	
	\begin{figure}[H]
		\centering
		\includegraphics[width=1.0\textwidth]{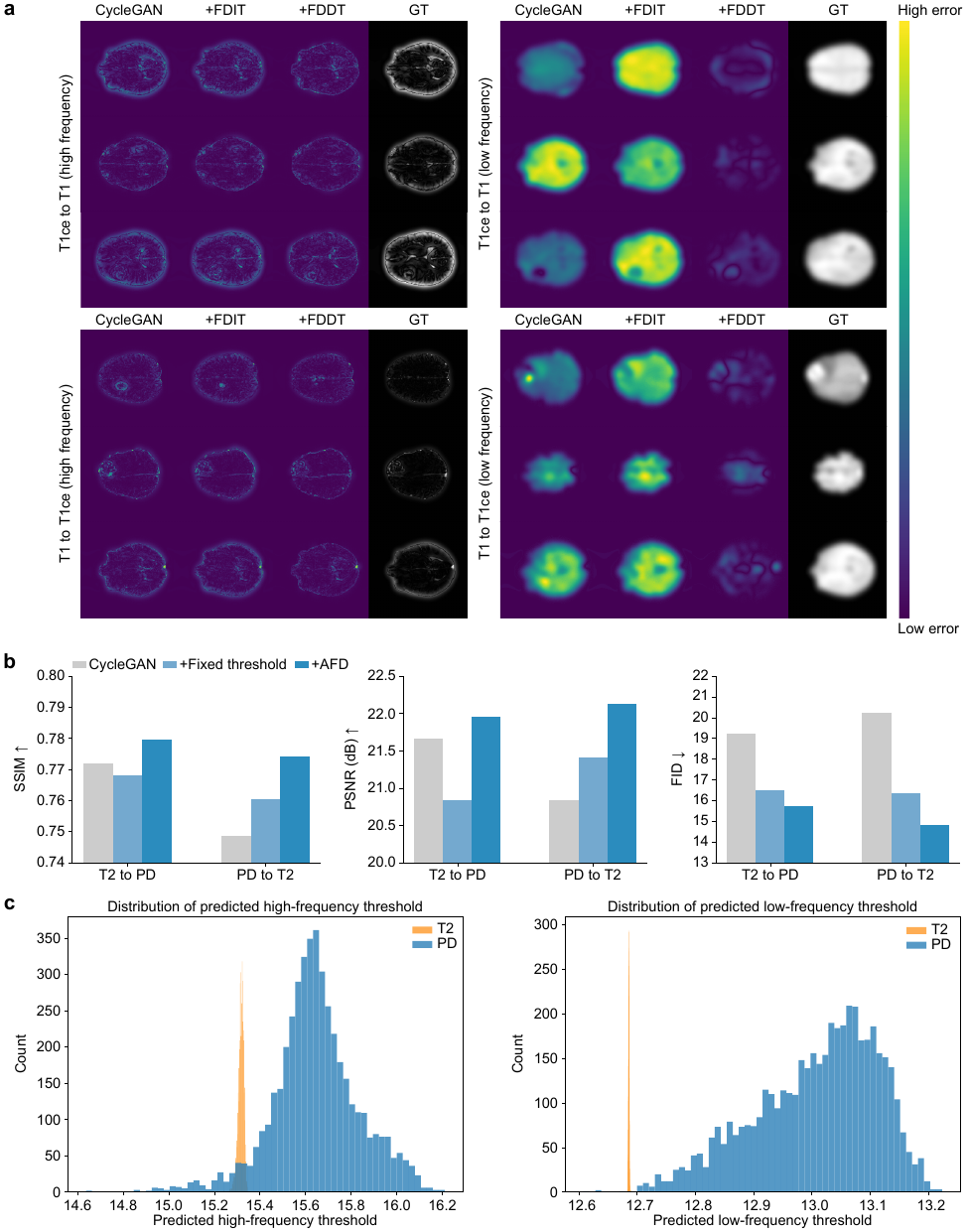}
		\caption{
			\textbf{Adaptive frequency decomposition.}
			\textbf{(a)} Pixelwise error maps of the high-frequency (left) and low-frequency (right) components for bidirectional T1\(\leftrightarrow\)T1ce translation on the held-out 219-case official BraTS 2021 validation cohort (T1ce\(\rightarrow\)T1, top; T1\(\rightarrow\)T1ce, bottom). Each row shows a representative axial slice from a different case. For CycleGAN, CycleGAN+FDIT and CycleGAN+FDDT, maps show the absolute difference between the corresponding frequency component of the synthesized image and that of the ground truth, on a shared colour scale (dark, low error; bright, high error) common to all panels. GT columns show the ground-truth frequency component in grayscale for reference.
			\textbf{(b)} SSIM, PSNR and FID for the base CycleGAN translator, fixed-threshold FDDT and FDDT with AFD on the 88-case IXI test cohort for T2\(\rightarrow\)PD and PD\(\rightarrow\)T2 translation. Bars show cohort means for SSIM and PSNR, which were calculated at the case level; FID was calculated at the dataset level.
			\textbf{(c)} Distributions of the sample-specific high- and low-frequency cutoffs predicted for the IXI T2 and PD images. Case-level SSIM and PSNR were compared using paired Wilcoxon signed-rank tests with Holm correction across the eight comparisons. Supplementary Table~\ref{tab:afd_sig_appendix} reports mean paired differences to align with the bars and separately reports median paired differences with 95\% percentile-bootstrap CIs obtained using 5,000 resamples. FID was reported descriptively.
		}
		\label{fig3}
	\end{figure}
	
	\begin{figure}[H]
		\centering
		\includegraphics[width=1.0\textwidth]{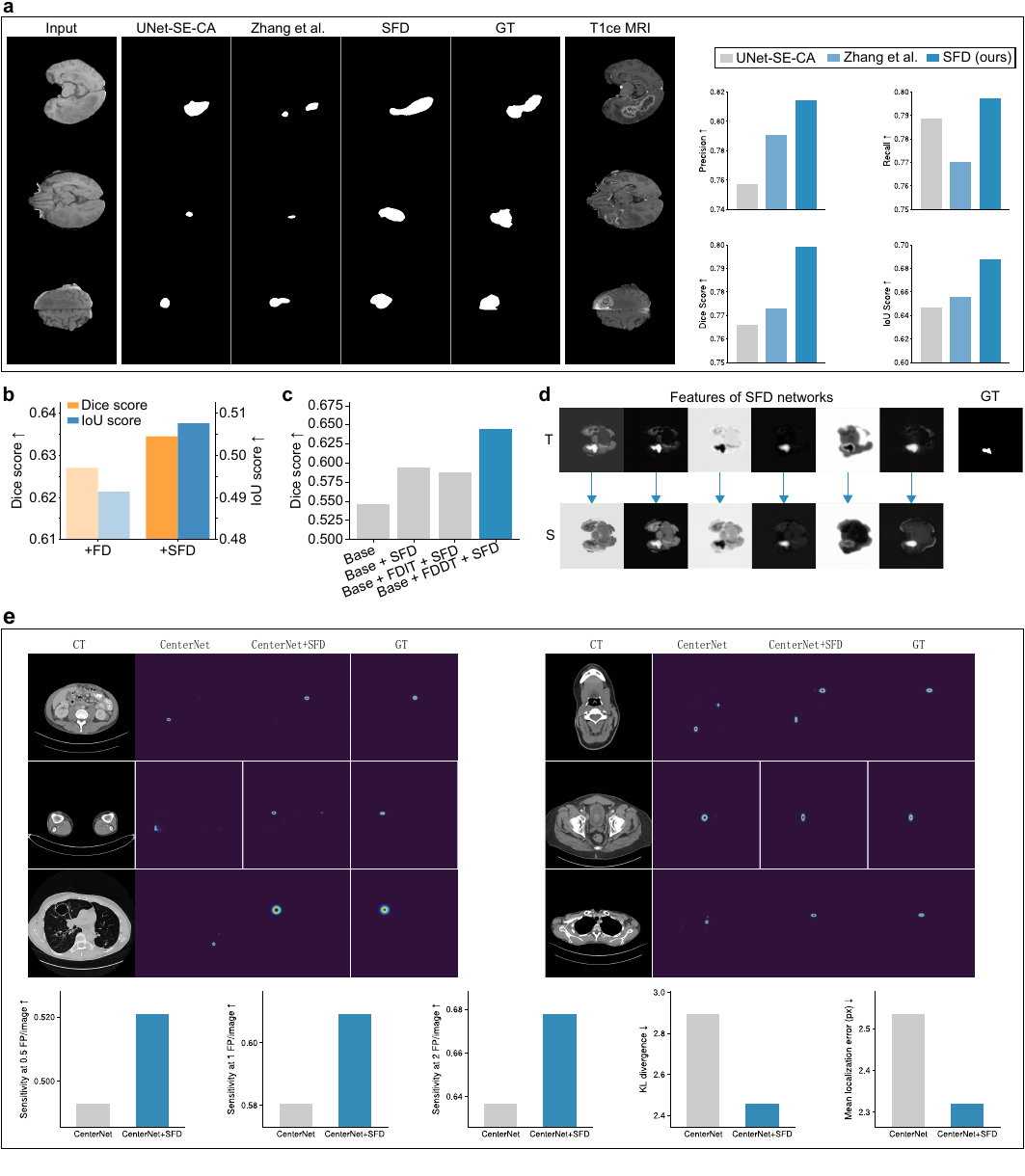}
		\caption{
			\textbf{Selective feature distillation improves tumour-containing-slice segmentation and lesion centre-point detection.}
			\textbf{(a)} Representative input images, predicted whole-tumour binary masks, ground-truth masks and T1ce MRI, together with case-level precision, recall, Dice and IoU on the 249-case held-out BraTS 2021 test partition. Predicted probabilities were thresholded at 0.5, and the metrics were calculated in three dimensions over the selected tumour-containing slices of each case.
			\textbf{(b)} Comparison of ungated full feature distillation (FD) and selective feature distillation (SFD) using the independently trained 80/16/16 ablation split.
			\textbf{(c)} Ablation of the base model and its combinations with SFD, FDIT and FDDT on the same 16-case test partition.
			\textbf{(d)} Representative student (S) and teacher (T) feature maps during SFD training, shown alongside the ground-truth mask.
			\textbf{(e)} Representative DeepLesion CT images and detection heatmaps, together with patient-wise macro-averaged sensitivity at FP budgets of 0.5, 1 and 2 false positives per image, KL divergence and conditional mean localization error for CenterNet with and without SFD. Sensitivity and KL divergence were evaluated in all 663 test patients; localization error was evaluated in the 528 patients with valid matched detections for both methods. Raw and Holm-adjusted results for panels \textbf{(a)} and \textbf{(e)} are reported in Supplementary Tables~\ref{tab:sfd_seg_sig_appendix} and~\ref{tab:sfd_det_sig_appendix}, respectively. Bars show case-level means in (a--c) and patient-level means in (e).
		}
		\label{fig4}
	\end{figure}
	
	\begin{figure}[H]
		\centering
		\includegraphics[width=1.0\textwidth]{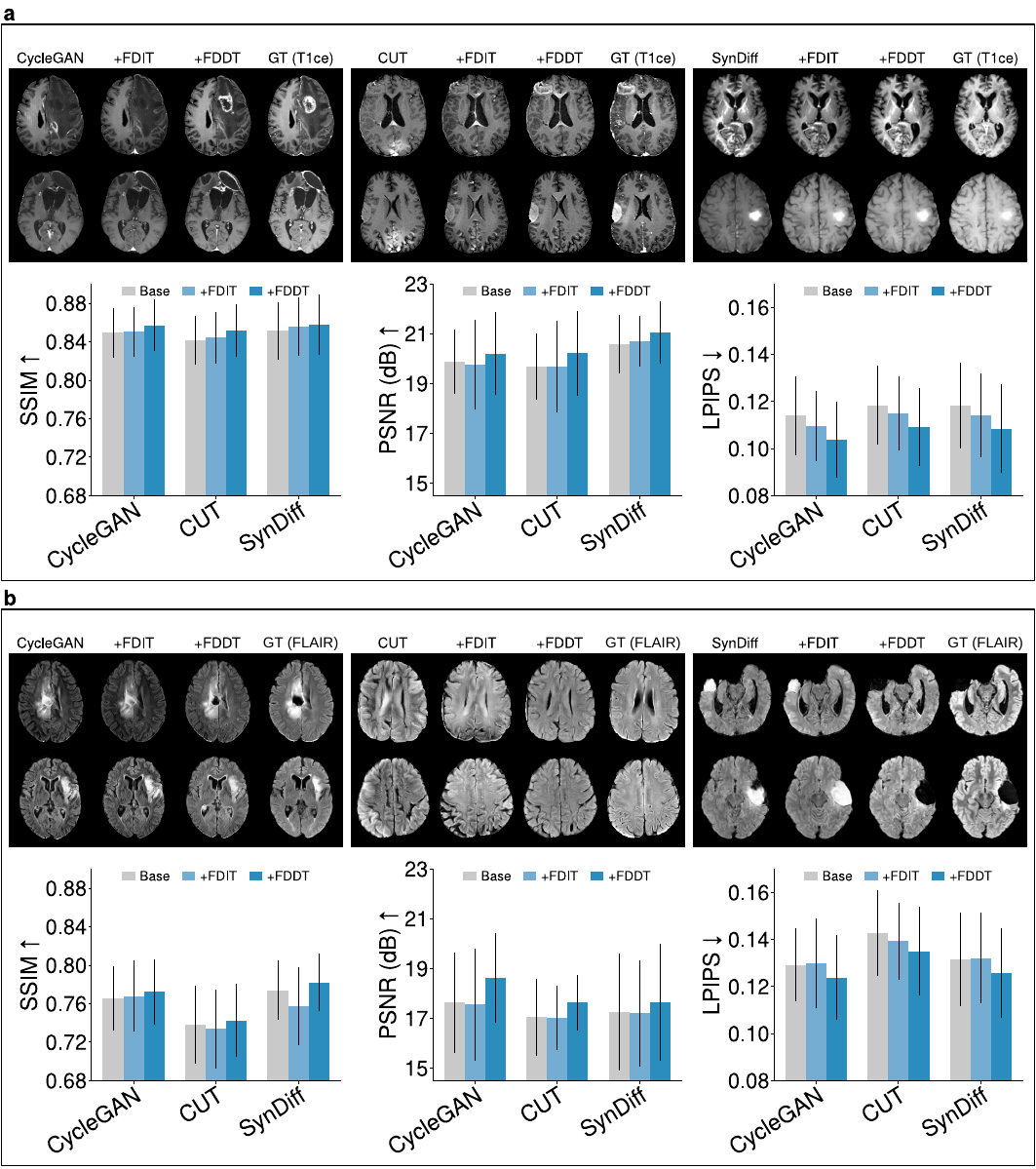}
		\caption{
			\textbf{External validation of image translation on BraTS 2024.}
			BraTS 2021-trained translators were evaluated without retraining on 87 patients from BraTS 2024, comprising 187 longitudinal post-treatment MRI examinations.
			\textbf{(a)} Representative T1\(\rightarrow\)T1ce translations and patient-level SSIM, PSNR and LPIPS.
			\textbf{(b)} Representative T2\(\rightarrow\)FLAIR translations and the corresponding patient-level metrics. For patients with multiple examinations, each metric was first calculated per examination and then averaged within patient, such that each patient contributed one value to the cohort-level summary and statistical analysis. GT denotes the paired target-modality image used only for evaluation. Bars show means across the 87 patients, and error bars show standard deviations (SDs) across patients. Raw and Holm-adjusted results are reported in Supplementary Table~\ref{tab:brats2024_sig_appendix}.
		}
		\label{fig5}
	\end{figure}
	
	\section*{Discussion}\label{sec:discussion}
	
	\par This study links medical image synthesis with downstream task learning through adaptive frequency consistency and selective teacher-feature transfer. Under the evaluated settings, the framework improved translation fidelity, brain-tumour segmentation on tumour-containing slices and lesion centre-point detection. FDDT improved translation metrics across CycleGAN, CUT and SynDiff without changing the inference architecture of the corresponding translator \cite{cyclegan,cut,syndiff}. Results on BraTS 2021, SynthRAD2025 and IXI, together with external translation evaluation on BraTS 2024, support applicability across the evaluated model families, modality pairs and data distributions. The downstream experiments demonstrate task-level gains, whereas the separate segmentation ablation provides the direct comparison between gated and ungated feature distillation.
	
	\par The core of FDDT is to apply frequency-consistency objectives to both low- and high-frequency components during translation. Because the frequency pathways share the same generator, this design encourages the translator to preserve information at both global-structure and fine-detail scales without introducing separate frequency-specific generator parameters \cite{gonzalez2009digital,fdit}. AFD replaces fixed cutoffs with sample-level learnable thresholds, enabling adaptive band selection that reflects sample-specific spatial and spectral characteristics. Empirically, these choices improve SSIM and PSNR while reducing LPIPS or FID, depending on the experiment \cite{wang2004image,huynh2008scope,lpips,heusel2017gans}. The frequency-component visualizations are consistent with low-frequency information contributing to global structure and high-frequency information contributing to fine image detail \cite{li2023zero,gao2024frequency}. SFD formulates feature transfer as an explicit per-sample comparison of task-specific risks: distillation is activated only when the teacher outperforms the student by the predefined margin, and otherwise the student is optimized without the feature-distillation term \cite{hinton2015distilling,romero2015fitnets}. This risk comparison uses task labels during training and does not introduce an inference-time gate. Unlike prediction rejection in selective prediction \cite{geifman2019selectivenet}, this gate controls feature distillation: it does not exclude the synthetic input from the student or guarantee that every transferred representation is beneficial. In the DeepLesion implementation, the comparison combines heatmap reconstruction, distributional agreement and peak-localization error. Relative to ungated full distillation in the segmentation ablation, SFD improved Dice and IoU; relative to the base CenterNet detector on DeepLesion, it improved sensitivity while reducing heatmap KL divergence and conditional localization error \cite{unet,centernet}. The additional training cost arises mainly from the teacher computation path and feature alignment. At deployment, the teacher and distillation branch are removed, whereas \(G\) and the translator-conditioned student are retained.
	
	\par The external evaluation from the preoperative BraTS 2021 cohort to the longitudinal post-treatment BraTS 2024 cohort showed that the translation improvements associated with FDDT persisted under a clinically relevant dataset shift without retraining \cite{brats2024_2}. The external evidence is limited to translation fidelity; downstream performance on post-treatment MRI was not evaluated. From a systems perspective, the additional training overhead arises from frequency-band decomposition, adaptive cutoff prediction and the teacher computation path. These training-only components have different deployment implications: FDDT/AFD inference remains equivalent to that of the base translator, whereas SFD inference consists of \(G\) followed by the student and should therefore be compared with the same translator-conditioned student pipeline rather than with an \(X\)-only task model. Accordingly, FDDT/AFD can be viewed as a training-time frequency-consistency regularizer that adds no translator inference cost, whereas SFD provides risk-gated feature distillation while retaining the deployment cost of the translator--student pipeline.
	
	\par Several limitations define the scope of the present evidence. The current frequency decomposition uses the DFT with isotropic Gaussian masks to construct independently parameterized low- and high-frequency representations, but it does not explicitly model spatially localized or direction-specific frequency patterns \cite{gonzalez2009digital}. Multi-resolution decompositions, wavelets or learnable filter banks could extend AFD while retaining its sample-adaptive cutoff prediction \cite{mallat2002theory}. SFD currently uses a hard gate derived from supervised task-specific risk; future extensions could incorporate uncertainty estimates or continuous weighting to further refine feature transfer. A limitation of the BraTS segmentation experiment is that evaluation was restricted to tumour-containing slices; therefore, false positives on tumour-negative slices were not assessed, and the reported results should not be interpreted as standard full-volume BraTS segmentation performance. Extending the framework to volumetric and spatiotemporal data, additional anatomical applications and downstream tasks would further examine its breadth. Although the present experiments span GAN- and diffusion-based translators, extensions to autoregressive and flow-based generators remain to be evaluated \cite{ddpm,ddim,ldm,fm}. The 16-case segmentation ablation also limits the evidence specifically isolating the gate from ungated distillation. The reported case- or patient-level CIs do not quantify variability across repeated model-training runs. Together, the results support synthetic modalities as task-oriented representations derived from acquired scans, not as independent patient measurements or substitutes for complementary clinical examinations. By removing the auxiliary frequency, AFD, teacher and distillation components after training, the framework improves translation fidelity and downstream task performance while limiting deployment cost to the base translator or the translator--student pipeline, respectively.
	
	\section*{Methods}\label{sec:method}
	
	\subsection*{Datasets and ethical considerations}
	
	\par We used five public, de-identified datasets spanning multiple centres, imaging protocols, anatomical regions and imaging modalities.
	
	\par\textbf{BraTS 2021.}
	We used co-registered, multimodal preoperative brain MRI examinations containing T1, T1ce, T2 and FLAIR images \cite{brats2021_1,brats2021_2,brats2021_3}. The official release contains 1,251 training cases with segmentation labels and 219 validation cases without publicly released segmentation labels.
	
	For the standalone image-translation experiments in Fig.~\ref{fig2}, the 1,251-case official BraTS 2021 training cohort was partitioned at the case level into 1,100 training cases and 151 internal validation cases. Translation models were optimized under an unpaired protocol using only the 1,100-case training partition, while the 151-case internal validation partition was used exclusively for predeclared checkpoint selection. After the final checkpoint for each model was locked, the 219-case official BraTS 2021 validation cohort was accessed once as a held-out final evaluation cohort. Source--target correspondence was not provided to the translation objective during training, and the paired target-modality images in the held-out cohort were used only for final evaluation.
	
	For the main downstream segmentation experiment in Fig.~\ref{fig4}a, the 1,251 labelled cases were partitioned at the case level into 900 training, 102 validation and 249 test cases. A separate segmentation-specific translator was pretrained in an initial translation stage using only the 900 training cases. The same 900-case partition was used for downstream model optimization, the 102-case partition was used only for model selection, and the 249-case partition was reserved exclusively for final testing.
	
	The segmentation ablations in Fig.~\ref{fig4}b,c used a separately defined 80/16/16 case-level training/validation/test split. The corresponding translator for each ablation setting was pretrained using only its 80-case training partition. The validation and test cases were not used during translator pretraining or downstream model optimization.
	
	\par\textbf{SynthRAD2025.}
	SynthRAD2025 provides paired CT and MR imaging from multiple anatomical regions and institutions \cite{synthrad2025}. We used the public Task 1 training archive from the official Zenodo release (\url{https://doi.org/10.5281/zenodo.14918213}), comprising 513 MRI--CT cases from Centers A, B and C under the CC BY-NC 4.0 license. Center E contributes no Task 1 training cases, and the additional 65 Center D cases distributed separately through the Grand Challenge platform were not included. We created non-overlapping patient-level partitions of 360 training cases, 50 validation cases and 103 test cases and used CT as the source modality and MR as the target modality. Sparse inter-slice sampling yielded 9,963 two-dimensional images per modality. Patient pairing was preserved only for defining patient-level partitions and calculating paired test-set SSIM and PSNR; source--target correspondence was not provided to the translation objective during training.
	
	\par\textbf{DeepLesion.}
	After excluding 34 key slices flagged as possibly noisy in the official metadata, the retained DeepLesion cohort contained 32,086 images from 4,422 patients. Following the official patient-level partition, 3,057/702/663 patients were used for training, validation and testing, respectively, with no patient overlap across subsets. Images and target heatmaps were resized to \(512\times512\) pixels. Lesion centres derived from the RECIST annotations were encoded as peaks in the target heatmaps for centre-point detection. The synthetic-modality generation, translator pretraining and fine-tuning, and ground-truth heatmap construction procedures are detailed in Supplementary Methods \S\ref{supp:sfd-objectives}.
	
	For the DeepLesion SFD experiment, the CT-to-MR translator was initialized using weights pretrained on the SynthRAD2025 training partition, with CT as the source domain and MR as the target domain. During downstream SFD training, the translator was further fine-tuned using the SynthRAD2025 training data alongside detector training on the DeepLesion training partition. The DeepLesion validation and test partitions were excluded from translator updates and downstream model optimization.
	
	\par\textbf{IXI (Information eXtraction from Images).}
	The IXI dataset comprises structural T1, T2 and PD brain MRI from 581 healthy participants acquired at three institutions in London, UK \cite{ixi_dataset}. Following image registration and quality control, 509 participants were retained, yielding 5,548 two-dimensional images each for T2 and PD. The patient-level training/validation/test split was 400/21/88 cases.
	
	\par All experiments kept patients or participants disjoint across their respective training, validation and test partitions.
	
	\par\textbf{External validation (BraTS 2024).}
	The BraTS 2021-trained translators were evaluated without retraining on 87 patients from the BraTS 2024 validation data, comprising 187 longitudinal post-treatment MRI examinations \cite{brats2024_2}. The same spatial and intensity preprocessing used for BraTS 2021 was applied. Examination-level metrics were averaged within patient across available time points, so that each patient contributed one value to the statistical analysis. BraTS 2024 was used exclusively for external evaluation.
	
	\par All datasets were publicly available and de-identified before release. This study comprised secondary analyses of de-identified data and was conducted in accordance with the original dataset licenses and applicable data-use requirements \cite{brats2021_1,synthrad2025,deeplesion,ixi_dataset,brats2024_2}.
	
	\subsection*{FDDT architecture and loss functions}
	
	\par Modern image-to-image translation builds on generative adversarial networks and diffusion models. Building on our initial GAN-based FDDT formulation \cite{wang2023fddt}, the present framework augments a base translator from either paradigm with training-time frequency decomposition and self-consistency constraints. Let \(X\in\mathcal{X}\subset\mathbb{R}^{H\times W\times C}\) denote an input image and let \(\widehat{X}_{c}\in\mathbb{C}^{H\times W\times C}\) denote its centered Fourier spectrum.
	
	\par We first compute the channel-wise DFT directly from \(X\) and shift the zero-frequency component to the center of the spectrum:
	\[
	\widehat{X}_{c}
	=
	\operatorname{fftshift}\!\left(\mathrm{DFT}(X)\right).
	\]
	The full summation form and normalization convention are provided in Supplementary Eq.~\eqref{eq:app-dft}.
	
	We then apply isotropic frequency masks to obtain low- and high-frequency bands:
	\begin{equation}
		\begin{aligned}
			\widehat{X}_L(u,v)&=M_L(u,v;D_l)\widehat{X}_{c}(u,v),\\
			\widehat{X}_H(u,v)&=M_H(u,v;D_h)\widehat{X}_{c}(u,v),
		\end{aligned}
	\end{equation}
	where \(\widehat{X}_L(u,v)\) and \(\widehat{X}_H(u,v)\) are the centered low- and high-frequency spectra; \(D_l>0\) and \(D_h>0\) are independently predicted, sample-adaptive bandwidth parameters; and multiplication is element-wise and applied independently to each channel. Defining centered coordinates
	\[
	\bar u=u-\left\lfloor\frac{H}{2}\right\rfloor,
	\qquad
	\bar v=v-\left\lfloor\frac{W}{2}\right\rfloor,
	\]
	the Gaussian masks are
	\begin{equation}
		\begin{aligned}
			M_L(u,v;D_l)&=\exp\!\left(-\frac{\bar u^2+\bar v^2}{2D_l^2}\right),\\
			M_H(u,v;D_h)&=1-\exp\!\left(-\frac{\bar u^2+\bar v^2}{2D_h^2}\right).
		\end{aligned}
	\end{equation}
	Because \(D_l\) and \(D_h\) are predicted by independent regression heads, \(M_L(u,v;D_l)+M_H(u,v;D_h)\) is not constrained to equal one. Accordingly, \(X_L\) and \(X_H\) are independently parameterized frequency-band representations that may overlap or leave a gap in frequency coverage; they are not required to form a complementary decomposition or to reconstruct \(X\) exactly.
	
	The spatial representations are obtained by undoing the frequency shift before the inverse DFT:
	\[
	X_L=\operatorname{Re}\!\left\{\mathrm{IDFT}\!\left(\operatorname{ifftshift}(\widehat{X}_L)\right)\right\},
	\qquad
	X_H=\operatorname{Re}\!\left\{\mathrm{IDFT}\!\left(\operatorname{ifftshift}(\widehat{X}_H)\right)\right\}.
	\]
	The real component of each inverse transform is used, preserving the sign of the spatial-frequency response; residual imaginary components caused by finite-precision arithmetic are discarded. The expanded inverse-transform expressions are provided in Supplementary Eq.~\eqref{eq:app-idft}.
	
	\par FDDT enforces self-consistency between the translation of each frequency-band input and the corresponding band extracted from the full-image translation. Denoting the low- and high-frequency operators by \((\cdot)_L\) and \((\cdot)_H\), respectively, the frequency-consistency loss is
	\begin{equation}
		\mathcal{L}_{\mathrm{freq}}(G)=
		\mathbb{E}_{X\sim\mathcal{X}}
		\left[
		\left\|G(X_L)-\left(G(X)\right)_L\right\|_1
		+
		\left\|G(X_H)-\left(G(X)\right)_H\right\|_1
		\right],
	\end{equation}
	where \(G\) denotes the translation mapping implemented by the selected GAN- or diffusion-based backbone, and \(\|\cdot\|_1\) is the element-wise \(\ell_1\) norm. The band operators applied to \(X\) and \(G(X)\) use the sample-specific cutoffs predicted for the corresponding image; gradient handling for the input and output paths is described in the AFD section below.
	
	The total training objective is
	\begin{equation}
		\mathcal{L}_{\mathrm{FDDT}}
		=
		\mathcal{L}_{\mathrm{base}}
		+
		\lambda_{\mathrm{freq}}\mathcal{L}_{\mathrm{freq}},
	\end{equation}
	where \(\mathcal{L}_{\mathrm{base}}\) denotes the backbone-specific translation objective and \(\lambda_{\mathrm{freq}}>0\) controls the frequency-consistency term. All frequency-domain operations and AFD predictions are applied exclusively during training; inference remains identical to that of the base translator.
	
	\subsection*{AFD: Adaptive frequency-domain decomposition and cutoff regression}
	
	\par AFD is a separately parameterized module that is jointly optimized with generator \(G\) through \(\mathcal{L}_{\mathrm{FDDT}}\). A single AFD instance is shared across the real-input and generated-output paths, so the same cut-off predictor \(A_{\psi}(\cdot)\) is applied to \(X\) and \(G(X)\).
	
	\par The input is a normalized single-channel image \(X\); the outputs are two cutoff parameters constrained to predefined ranges,
	\[
	D_h\in[D_{\min}^{H},D_{\max}^{H}],
	\qquad
	D_l\in[D_{\min}^{L},D_{\max}^{L}],
	\]
	where \(D_h=D_h(X)\) and \(D_l=D_l(X)\) are sample-specific frequency cutoffs. AFD predicts the cutoffs from a grayscale input; for multi-channel images, the resulting cutoff pair is shared across channels.
	
	\par We define magnitude and phase features as \(M(X)\) and \(\Phi(X)\). Their full definitions are provided in Supplementary Eq.~\eqref{eq:freq-feat}. We construct two feature streams:
	\begin{equation}\label{eq:streams}
		\begin{aligned}
			S(X)&=\mathrm{CNN}_s(X),\\
			F(X)&=\mathrm{CNN}_f\!\left([M(X),\Phi(X)]\right),
		\end{aligned}
	\end{equation}
	where \(S(X)\) is the spatial-stream feature map, \(F(X)\) is the frequency-stream feature map, and \([\cdot,\cdot]\) denotes channel-wise concatenation.
	
	\par We flatten the feature maps into token sequences, using the spatial-stream tokens as queries \(Q\) and the frequency-stream tokens as keys \(K\) and values \(V\), and apply cross-attention from the spatial stream to the frequency stream. The explicit tokenization and multi-head attention definitions are provided in Supplementary Eqs.~\eqref{eq:qkv}--\eqref{eq:mha}. The attention output is reshaped and fused with the spatial stream:
	\begin{equation}\label{eq:fuse}
		U(X)=\mathrm{Fuse}\!\left([S(X),A(X)]\right).
	\end{equation}
	
	Two independent regression heads predict normalized cutoff scores:
	\begin{equation}\label{eq:heads}
		\begin{aligned}
			\hat{d}_h(X)&=\sigma\!\left(w_h^\top\mathrm{vec}(U(X))+b_h\right),\\
			\hat{d}_l(X)&=\sigma\!\left(w_l^\top\mathrm{vec}(U(X))+b_l\right),
		\end{aligned}
	\end{equation}
	where \(\sigma(t)=1/(1+e^{-t})\), and \(\mathrm{vec}(U(X))\) denotes direct flattening of all channel and spatial dimensions of \(U(X)\) without global or adaptive pooling.
	
	The scores are mapped to the predefined cutoff ranges:
	\begin{equation}\label{eq:minmax}
		\begin{aligned}
			D_h(X)&=D_{\min}^{H}
			+\left(D_{\max}^{H}-D_{\min}^{H}\right)\hat d_h(X),\\
			D_l(X)&=D_{\min}^{L}
			+\left(D_{\max}^{L}-D_{\min}^{L}\right)\hat d_l(X).
		\end{aligned}
	\end{equation}
	Together, the two regression heads define \(A_{\psi}(X)=(D_h(X),D_l(X))\).
	
	When the predicted cutoffs are used to parameterize masks for real input images, we apply a stop-gradient operator:
	\begin{equation}\label{eq:sg}
		\begin{aligned}
			D_h^{\mathrm{sg}}(X)&=\mathrm{sg}\!\left[D_h(X)\right],\\
			D_l^{\mathrm{sg}}(X)&=\mathrm{sg}\!\left[D_l(X)\right].
		\end{aligned}
	\end{equation}
	where \(\mathrm{sg}(z)=z\) in the forward pass and \(\partial\mathrm{sg}(z)/\partial z=0\). For the generated-output path, \(A_{\psi}(G(X))\) remains differentiable, allowing gradients to propagate through AFD and \(G\). For the real-input path, the detached cutoffs \(D_h^{\mathrm{sg}}(X)\) and \(D_l^{\mathrm{sg}}(X)\) are used. This asymmetric gradient flow blocks gradients through the real-input cutoff predictions while retaining gradients through the generated-output cutoff path.
	
	The AFD module and FDDT frequency-consistency objectives were used exclusively during translator pretraining and were disabled during downstream SFD fine-tuning. Only the pretrained base translator was retained for subsequent joint optimization with the downstream task networks.
	
	\subsection*{SFD: Selective feature distillation}
	
	\par SFD uses the image \(\tilde{Y}\) generated by the FDDT-trained translator as an auxiliary condition. For the BraTS 2021 segmentation experiments, translator G was pretrained using only the corresponding segmentation training partition. For DeepLesion centre-point detection, the CT-to-MR translator was initialized using weights pretrained on the SynthRAD2025 training partition and subsequently fine-tuned with continued use of SynthRAD2025 training data during downstream SFD training, while the detection networks were trained on the DeepLesion training partition. The corresponding downstream validation and test partitions were excluded from translator optimization and downstream task-network training. Further details of the DeepLesion data flow are provided in Supplementary Methods \S\ref{supp:sfd-objectives}. Translator \(G\) is then used to generate \(\tilde{Y}\) from \(X\). The student receives the channel-wise concatenated input \([X,\tilde{Y}]\), whereas the teacher receives the duplicated translated input \([\tilde{Y},\tilde{Y}]\), preserving the same input dimensionality. During SFD training, the pretrained \(G\), student, and teacher are jointly fine-tuned. The teacher has its own downstream task loss and is used to compute the task-specific risk gate and provide stop-gradient feature targets:
	\begin{equation}\label{eq:sfd-input}
		\begin{aligned}
			\tilde Y&=G(X),\\
			U_S&=[X,\tilde Y],\\
			U_T&=[\tilde Y,\tilde Y],\\
			p_S&=s_\theta(U_S),\qquad p_T=t_\eta(U_T).
		\end{aligned}
	\end{equation}
	where \(s_\theta\) and \(t_\eta\) denote the student and teacher networks. For the compact notation below, \(f\in\{s_\theta,t_\eta\}\) and \(U\) denotes the corresponding network input: \(U=U_S\) when \(f=s_\theta\), and \(U=U_T\) when \(f=t_\eta\). The target variable \(\tau\) denotes the segmentation mask \(m\) or the ground-truth detection heatmap \(H^{\mathrm{gt}}\), as appropriate.
	
	\par The downstream task loss \(\mathcal{L}_{\mathrm{d}}\) is defined according to the task. For segmentation, we use a weighted combination of binary cross-entropy, focal, IoU, and Dice losses. For DeepLesion detection, the heatmap objective combines balanced BCE, distributional KL, peak-positive, hard-background, ranking, mass-calibration, and MSE terms. The corresponding weighted objectives and the detection gating risk are provided in Supplementary Eqs.~\eqref{eq:sfd-seg}--\eqref{eq:sfd-det-risk}:
	\begin{equation}\label{eq:sfd-obj}
		\mathcal{L}_{\mathrm{d}}(f;U,\tau)=
		\begin{cases}
			\mathcal{L}_{\mathrm{seg}}(f;U,m), & \text{segmentation},\\
			\mathcal{L}_{\mathrm{det}}(f;U,H^{\mathrm{gt}}), & \text{detection}.
		\end{cases}
	\end{equation}
	
	\par SFD applies feature distillation only when the student underperforms the teacher by more than a task-specific margin:
	\begin{equation}\label{eq:sfd-gate}
		g_i=
		\begin{cases}
			1, & R_i^T+m_{\mathrm{d}}<R_i^S,\\
			0, & \text{otherwise}.
		\end{cases}
	\end{equation}
	For segmentation, \(R_i^S\) and \(R_i^T\) are the per-sample segmentation objectives of the student and teacher, respectively, and \(m_{\mathrm{d}}=0\). For DeepLesion detection, \(R_i^S\) and \(R_i^T\) are calculated using the separately defined gating risk in Supplementary Eq.~\eqref{eq:sfd-det-risk}, which is distinct from the full detection training objective, and \(m_{\mathrm{d}}=0.02\). The resulting binary gate \(g_i\) is treated as non-differentiable during optimization.
	
	\par Using this gate, we align student features with stop-gradient teacher features:
	\begin{equation}\label{eq:sfd-distill}
		\mathcal{L}_{\mathrm{distill}}
		=
		\frac{1}{|\mathcal{S}|}\sum_{\ell\in\mathcal{S}}
		\frac{1}{B}
		\sum_{i=1}^{B}
		\frac{g_i}{C_\ell H_\ell W_\ell}
		\left\|
		F_{S,i}^{\ell}(U_S)
		-
		\mathrm{sg}\!\left[F_{T,i}^{\ell}(U_T)\right]
		\right\|_2^2.
	\end{equation}
	The layer-wise distillation losses are averaged across selected layers. Here \(F_{S,i}^{\ell}\) and \(F_{T,i}^{\ell}\) denote the student and teacher feature maps for sample \(i\) at a selected layer \(\ell\in\mathcal{S}\), and \(C_\ell\), \(H_\ell\), and \(W_\ell\) denote their channel and spatial dimensions. For DeepLesion detection, the spatially weighted and channel-normalized formulation in Supplementary Eq.~\eqref{eq:sfd-det-distill} is used.
	
	\par The overall SFD objective is
	\begin{equation}\label{eq:sfd-total}
		\mathcal{L}_{\mathrm{sfd}}
		=
		\mathcal{L}_{\mathrm{d}}^{S}
		+
		\mathcal{L}_{\mathrm{d}}^{T}
		+
		\lambda_{\mathrm{distill}}\mathcal{L}_{\mathrm{distill}}
		+
		\lambda_{\mathrm{GAN}}\mathcal{L}_{\mathrm{GAN}}.
	\end{equation}
	Here, \(\lambda_{\mathrm{distill}}\) is initialized as zero during the first 20 warm-up epochs and linearly increased over the subsequent 30 epochs to the final value \(\lambda_{\mathrm{distill}}=0.50\), \(\mathcal{L}_{\mathrm{GAN}}\) denotes the original adversarial GAN loss of the backbone translator retained during SFD fine-tuning, and its weight satisfies \(\lambda_{\mathrm{GAN}}=0.05\) in all downstream experiments. The translator, student, and teacher are jointly optimized during SFD training. The stop-gradient in Eq.~\eqref{eq:sfd-distill} blocks gradients from the feature-matching term into the teacher, while the teacher remains trainable through \(\mathcal{L}_{\mathrm{d}}^{T}\). At inference, only translator \(G\) and student \(s_\theta\) are retained; teacher \(t_\eta\) and the distillation branch are removed.
	
	\subsection*{Training and implementation details}
	
	\par Experiments were implemented using PyTorch v1.12.1 and CUDA v11.3 on NVIDIA TITAN RTX GPUs. Training batch sizes were 1 for image translation, 2 for segmentation, and 3 for DeepLesion detection. GAN-based translators and segmentation networks were optimized using Adam with an initial learning rate of \(10^{-4}\). Diffusion-based translators were optimized using AdamW with an initial learning rate of \(10^{-5}\). The DeepLesion models were optimized using AdamW with an initial learning rate of \(10^{-4}\), a weight decay of \(10^{-4}\), and a cosine-annealing learning-rate schedule over a maximum of 500 epochs.
	
	\subsection*{Baselines and comparisons}
	
	\par CycleGAN and CUT were used as GAN-based translation baselines, SynDiff as a diffusion-based baseline, and Frequency Domain Image Translation (FDIT) was integrated into each backbone as the frequency-domain comparison \cite{cyclegan,fdit,cut,syndiff}. Within each translation backbone, the unmodified, FDIT and FDDT variants used the same case-level partitions and modality-specific preprocessing. Source--target correspondence was withheld from the translation training objective; paired held-out images were used for evaluation. For segmentation, the proposed SFD model was compared with a U-Net equipped with squeeze-and-excitation and coordinate attention (UNet-SE-CA), and the cycle- and shape-consistency translation--segmentation method of Zhang et al. \cite{unet,senet,ca-attn,cvpr18}. For DeepLesion, the same U-Net-based CenterNet-style heatmap detector was trained and evaluated with and without SFD \cite{centernet}. Ungated full feature distillation (FD) was evaluated only in the independently trained segmentation ablation.
	
	\subsection*{Evaluation metrics and statistics}
	
	\par For translation, evaluation was performed at the independent case or patient level. BraTS 2021 and the external BraTS 2024 cohort were evaluated with structural similarity (SSIM), peak signal-to-noise ratio (PSNR), and learned perceptual image patch similarity (LPIPS) \cite{wang2004image,huynh2008scope,lpips}. SynthRAD2025 and the IXI ablation experiments were evaluated with SSIM, PSNR, and Fr\'echet Inception Distance (FID) \cite{heusel2017gans}. SSIM, PSNR and LPIPS were calculated against the corresponding target images when those metrics were used. Higher SSIM and PSNR indicate better performance, whereas lower LPIPS and FID are preferred. FID was computed at the dataset level and was not included in paired case-level hypothesis testing.
	
	\par For segmentation, the target was a single whole-tumour binary mask. Predicted probabilities were binarized at 0.5. For each case, the selected tumour-containing slices were assembled into a three-dimensional volume, and voxelwise true positives, false positives, and false negatives were accumulated over that volume before calculating case-level Precision, Recall, Dice, and IoU. Only slices with non-empty tumour annotations were included. Therefore, the reported segmentation results represent tumour-containing-slice evaluation rather than standard full-volume BraTS segmentation, and false positives on tumour-negative slices were not assessed. For DeepLesion FROC sensitivity evaluation, local maxima were extracted using \(9\times9\) non-maximum suppression. Candidate peaks with scores of at least \(10^{-4}\) were retained up to a maximum of 50 predictions per image and matched to ground-truth centres within a 15-pixel radius. For each patient, candidate peaks from all included images were pooled and ranked by confidence, and sensitivity was calculated at false-positive budgets corresponding to 0.5, 1, and 2 false positives per image. Patient-level sensitivities were then macro-averaged across patients. We additionally report heatmap KL divergence and mean matched localization error in pixels, which were computed at the image level and subsequently aggregated by patient. The three sensitivity metrics and KL divergence included all 663 test patients. Mean localization error was calculated only for the 528 patients with valid matched detections for both methods; undetected lesions were excluded from this conditional localization metric.
	
	\par All statistical analyses used one scalar value per independent patient or case. For BraTS 2024, examination-level metrics were averaged within patient across available time points before statistical testing. The Fig.~3b IXI ablations used paired Wilcoxon signed-rank tests. Mean paired differences were reported to align the effect estimates with the cohort-mean bars, while median paired differences were summarized separately with 95\% percentile-bootstrap CIs (5,000 resamples; seed 42). The available paired metrics for the other translation experiments, segmentation, and patient-aggregated DeepLesion analyses used paired \(t\)-tests with mean paired differences and 95\% \(t\)-intervals. All paired \(t\)-tests and Wilcoxon signed-rank tests were two-sided.
	
	\par Holm correction was applied separately to the complete set of comparisons in each supplementary table, defining six families with 36, 12, 8, 36, 8, and 5 tests for Tables~\ref{tab:stat_fig2a}--\ref{tab:sfd_det_sig_appendix}, respectively. Statistical significance was defined as Holm-adjusted \(p<0.05\). All CIs remain pointwise and unadjusted for multiplicity. Statistical results are provided in Supplementary Tables~\ref{tab:stat_fig2a}--\ref{tab:sfd_det_sig_appendix}. Thus, family-wise error control applies within each table, not jointly across all six tables.
	
	\section*{Data availability}\label{sec:data-availability}
	
	The datasets used in this study are publicly available from their official repositories, subject to their respective data-use terms. The SynthRAD2025 Task 1 training archive used here was obtained from Zenodo (\url{https://doi.org/10.5281/zenodo.14918213}); the separately distributed Center D data were not used. BraTS 2021 is available through Synapse (\url{https://www.synapse.org/Synapse:syn25829067/wiki/610863}), BraTS 2024 through the official challenge site (\url{https://www.synapse.org/brats2024}), DeepLesion through the NIH Clinical Center repository (\url{https://nihcc.app.box.com/v/DeepLesion}), and IXI through the IXI Project site (\url{https://brain-development.org/ixi-dataset/}).
	
	\section*{Code availability}\label{sec:code-availability}
	
	The main training and inference code is publicly available via GitHub at \url{https://github.com/fancycodeeer/From-Generation-to-Diagnosis}.
	
	\section*{Funding}
	
	This work was supported by the National Key R\&D Program of China (grant no. 2022YFC2402103), the National Natural Science Foundation of China (grant no. 31870942), the Beijing Natural Science Foundation (grant no. L259081), the Peking University Clinical Medicine Plus X--Young Scholars Project (grant nos. PKU2020LCXQ017 and PKU2021LCXQ028), and the PKU--Baidu Fund (grant no. 2020BD039).
	
	\section*{Author contributions}
	
	Z.W. conceived the study, developed the methodology and software, performed the experiments and statistical analyses, prepared the figures, and wrote the original draft. J.Z. contributed to the experiments, data analysis, and validation. X.P. contributed to clinical interpretation and validation. K.Z., X.D., and J.L. supervised the study, contributed to study design and interpretation of the results, and critically revised the manuscript. All authors reviewed and approved the final manuscript.
	
	\section*{Competing interests}
	
	The authors declare no competing interests.
	
	\bibliography{newref-revised}

	\clearpage
	\begingroup
	\setlength{\emergencystretch}{2em}
	\setcounter{page}{1}
	\renewcommand{\thepage}{S\arabic{page}}
	\setcounter{section}{0}
	\setcounter{subsection}{0}
	\setcounter{equation}{0}
	\setcounter{table}{0}
	\renewcommand{\thesection}{\arabic{section}}
	\renewcommand{\thesubsection}{\thesection.\arabic{subsection}}
	\renewcommand{\theequation}{S\arabic{equation}}
	\renewcommand{\thetable}{S\arabic{table}}
	\renewcommand{\theHsection}{supp.\arabic{section}}
	\renewcommand{\theHsubsection}{supp.\arabic{section}.\arabic{subsection}}
	\renewcommand{\theHequation}{supp.\arabic{equation}}
	\renewcommand{\theHtable}{supp.\arabic{table}}
	\renewcommand{\tablename}{Supplementary Table}
	\newcolumntype{S}[1]{>{\scriptsize\centering\arraybackslash}p{#1}}
	\newcolumntype{L}[1]{>{\scriptsize\raggedright\arraybackslash}p{#1}}
	
	\begin{center}
		{\Large\bfseries Supplementary Information\par}
		\vspace{1.2em}
		{\large\bfseries Frequency-consistent synthesis and selective transfer for medical image analysis\par}
		\vspace{1em}
		{Zhuhui Wang, Jianwei Zuo, Xin Peng, Kai Zhong, Xuliang Deng, and Jiajia Luo\par}
	\end{center}
	
	\vspace{1em}
	
	\noindent This Supplementary Information provides expanded formulations for frequency-domain decomposition, adaptive cutoff estimation, cross-attention fusion, and task-specific selective feature distillation (SFD) during supervised training. It then reports the paired statistical analyses supporting Figs.~\ref{fig2}--\ref{fig5} of the main text. Equation, table, and figure references follow the notation used in the main manuscript.
	
	\section{Supplementary Methods}
	\label{supp:methods}
	
	\subsection{Frequency-domain decomposition and adaptive cutoff estimation}
	\label{supp:frequency-decomposition}
	
	The discrete Fourier transform (DFT) is applied directly to the input image \(X\). We use the standard backward-normalization convention implemented by \texttt{torch.fft.fft2}: the forward transform is unnormalized and the inverse transform contains the factor \(1/(HW)\). The channel-wise unshifted and centered spectra are
	\begin{equation}\label{eq:app-dft}
		\begin{aligned}
			\widehat{X}(u,v)
			&=\mathrm{DFT}(X)(u,v)\\
			&=\sum_{h=0}^{H-1}\sum_{w=0}^{W-1}
			X(h,w)
			e^{-\,\mathrm{i}2\pi\left(\frac{uh}{H}+\frac{vw}{W}\right)},\\
			\widehat{X}_{c}
			&=\operatorname{fftshift}\!\left(\widehat{X}\right).
		\end{aligned}
	\end{equation}
	
	Here, \(\mathrm{i}=\sqrt{-1}\) denotes the imaginary unit. Defining the centered frequency coordinates
	\[
	\bar u=u-\left\lfloor\frac{H}{2}\right\rfloor,
	\qquad
	\bar v=v-\left\lfloor\frac{W}{2}\right\rfloor,
	\]
	the low- and high-frequency Gaussian masks and their corresponding centered masked spectra are
	\begin{equation}\label{eq:supp-masks}
		\begin{aligned}
			M_L(u,v;D_l)
			&=\exp\!\left(-\frac{\bar u^2+\bar v^2}{2D_l^2}\right),
			&\widehat{X}_L&=M_L\odot\widehat{X}_c,\\
			M_H(u,v;D_h)
			&=1-\exp\!\left(-\frac{\bar u^2+\bar v^2}{2D_h^2}\right),
			&\widehat{X}_H&=M_H\odot\widehat{X}_c.
		\end{aligned}
	\end{equation}
	The sample-specific bandwidths \(D_l\) and \(D_h\) are predicted independently by AFD. Therefore, \(M_L(u,v;D_l)+M_H(u,v;D_h)\) is not constrained to equal one. The resulting low- and high-frequency representations may overlap or leave a gap in frequency coverage; they are not required to form a complementary decomposition or to reconstruct \(X\) exactly.
	
	After undoing the spectral shift, the masked spectra are transformed back to the spatial domain. The spatial-band representations are obtained from the real components of the inverse transforms:
	\begin{equation}\label{eq:app-idft}
		\begin{aligned}
			X_L(h,w)
			&=\operatorname{Re}\!\left\{\frac{1}{HW}
			\sum_{u=0}^{H-1}\sum_{v=0}^{W-1}
			\operatorname{ifftshift}(\widehat{X}_L)(u,v)
			e^{\,\mathrm{i}2\pi\left(\frac{uh}{H}+\frac{vw}{W}\right)}\right\},\\
			X_H(h,w)
			&=\operatorname{Re}\!\left\{\frac{1}{HW}
			\sum_{u=0}^{H-1}\sum_{v=0}^{W-1}
			\operatorname{ifftshift}(\widehat{X}_H)(u,v)
			e^{\,\mathrm{i}2\pi\left(\frac{uh}{H}+\frac{vw}{W}\right)}\right\}.
		\end{aligned}
	\end{equation}
	Here, \(\operatorname{Re}\{\cdot\}\) denotes the element-wise real component. For a real input image, the Fourier spectrum is conjugate-symmetric. Multiplication by the real, even Gaussian frequency masks preserves this symmetry, so the inverse transform is real up to numerical round-off. Discarding the residual imaginary component preserves the sign of the spatial-frequency response.
	
	For adaptive cutoff estimation, AFD combines the spatial image with the log-magnitude and phase spectra
	\begin{equation}\label{eq:freq-feat}
		\begin{aligned}
			M(X)&=\log\!\left(
			\left|
			\operatorname{fftshift}(\mathrm{DFT}(X))
			\right|+1\right),\\
			\Phi(X)&=
			\angle\operatorname{fftshift}(\mathrm{DFT}(X)).
		\end{aligned}
	\end{equation}
	Here, \(M(X)\) and \(\Phi(X)\) denote the log-magnitude and phase spectra, respectively, and \(\log\) denotes the natural logarithm.
	
	\subsection{Cross-attention fusion in AFD}
	\label{supp:afd-attention}
	
	AFD forms a spatial stream \(S(X)\) and a frequency stream \(F(X)\) from the image and the concatenated log-magnitude and phase features, respectively. The feature maps are flattened into token sequences so that the spatial tokens query the frequency tokens:
	\begin{equation}\label{eq:qkv}
		\begin{aligned}
			Q&=\mathrm{flat}(S(X))\in\mathbb{R}^{L_s\times d},\\
			K&=\mathrm{flat}(F(X))\in\mathbb{R}^{L_f\times d},\\
			V&=K.
		\end{aligned}
	\end{equation}
	In Eq.~\eqref{eq:qkv}, \(L_s\) and \(L_f\) denote the numbers of spatial- and frequency-stream tokens, respectively, and \(d\) is the token embedding dimension. Multi-head cross-attention is computed as
	\begin{equation}\label{eq:mha}
		\mathrm{MHA}(Q,K,V)
		=
		\mathrm{Concat}_{h=1}^{n_h}
		\left[
		\mathrm{softmax}
		\left(
		\frac{QW_h^Q(KW_h^K)^\top}{\sqrt{d_h}}
		\right)
		VW_h^V
		\right]W^O.
	\end{equation}
	Here, \(n_h\) is the number of attention heads and \(d_h\) is the dimension of each head. The softmax operation is applied over the frequency-token dimension. The attention output is reshaped to the spatial feature layout and fused with \(S(X)\) to obtain the representation used by the two cutoff-regression heads described in the main text.
	
	\subsection{Task-specific SFD objectives}
	\label{supp:sfd-objectives}
	
	For brain-tumour segmentation, the downstream training objective is
	\begin{equation}\label{eq:sfd-seg}
		\mathcal{L}_{\mathrm{seg}}
		=
		0.25\mathcal{L}_{\mathrm{BCE}}
		+
		1.75\mathcal{L}_{\mathrm{Focal}}^{(\alpha=0.25,\gamma=5)}
		+
		\mathcal{L}_{\mathrm{IoU}}
		+
		\mathcal{L}_{\mathrm{Dice}}.
	\end{equation}
	The corresponding per-sample objective is used to calculate the student and teacher segmentation risks \(R_i^S\) and \(R_i^T\) in the hard gate.
	
	For DeepLesion centre-point detection, the CT-to-MR translator was initialized using weights pretrained on the SynthRAD2025 training partition (360 cases), with CT as the source modality and MR as the target modality. During downstream SFD training, the translator was further fine-tuned using the SynthRAD2025 training data, while the detection networks were optimized using the DeepLesion training partition. The student received both the original CT image and its synthetic MR counterpart, whereas the teacher received the synthetic MR image in both input channels. The DeepLesion dataset contained 32,086 CT slices from 4,422 patients after excluding annotations flagged as possibly noisy. Following the official patient-level partition, 22,478, 4,791, and 4,817 slices were assigned to training, validation, and testing, respectively, with no patient overlap. The validation and test partitions were excluded from model optimization.
	
	Ground-truth heatmaps were generated from the RECIST annotations provided in \texttt{DL\_info.csv}, comprising 32,700 lesion annotations across 32,086 unique CT slices. For each lesion, a rectangular bounding box was derived from the RECIST coordinates, and the lesion centre was defined as the geometric centre of the bounding box, \( (x_c,y_c)=((x_1+x_2)/2,(y_1+y_2)/2) \). Each lesion centre was represented by a two-dimensional anisotropic Gaussian, with standard deviations determined by the bounding-box width and height as \(\sigma_x=\operatorname{clip}(w/6,1,30)\) and \(\sigma_y=\operatorname{clip}(h/6,1,30)\), respectively. Gaussian responses were truncated at three standard deviations, with peak values set to 1. For slices containing multiple lesions, individual Gaussian responses were independently computed and combined into a single heatmap using a pixel-wise maximum operation. Heatmaps were initially generated at the corresponding image resolution without additional downsampling. During detector training, CT images and their corresponding heatmaps were synchronously resized to \(512\times512\) pixels to maintain spatial alignment.
	
	For DeepLesion centre-point detection, the downstream training objective is
	\begin{equation}\label{eq:sfd-det}
		\begin{aligned}
			\mathcal{L}_{\mathrm{det}}
			={}&
			\mathcal{L}_{\mathrm{balBCE}}
			+0.03\mathcal{L}_{\mathrm{KL}}
			+0.90\mathcal{L}_{\mathrm{peak}}
			+0.45\mathcal{L}_{\mathrm{hardBG}}\\
			&+0.30\mathcal{L}_{\mathrm{rank}}
			+0.02\mathcal{L}_{\mathrm{mass}}
			+0.05\mathcal{L}_{\mathrm{MSE}}.
		\end{aligned}
	\end{equation}
	
	Here, \(\mathcal{L}_{\mathrm{balBCE}}\) denotes the balanced binary cross-entropy loss for sparse heatmap prediction. Let \(z_p\in\mathbb{R}\) and \(y_p\in[0,1]\) denote the predicted logit and ground-truth heatmap value at pixel \(p\), respectively, and
	\[
	\ell_p
	=
	-y_p\log\sigma(z_p)
	-(1-y_p)\log(1-\sigma(z_p)).
	\]
	The balanced BCE is defined as
	\begin{equation}\label{eq:supp-bal-bce}
		\mathcal{L}_{\mathrm{balBCE}}
		=
		w_{\mathrm{fg}}
		\frac{\sum_p y_p^{\gamma_{\mathrm{fg}}}\ell_p}
		{\sum_p y_p^{\gamma_{\mathrm{fg}}}+\epsilon}
		+
		w_{\mathrm{bg}}
		\frac{\sum_p(1-y_p)^{\gamma_{\mathrm{bg}}}\ell_p}
		{\sum_p(1-y_p)^{\gamma_{\mathrm{bg}}}+\epsilon},
	\end{equation}
	where \(w_{\mathrm{fg}}=1.0\), \(w_{\mathrm{bg}}=0.25\), \(\gamma_{\mathrm{fg}}=1\), and \(\gamma_{\mathrm{bg}}=4\).
	
	The peak-positive loss focuses optimization on lesion-center regions. The peak core set \(\Omega_{\mathrm{peak}}\) is obtained from ground-truth heatmaps using a threshold of 0.7. If insufficient pixels are selected, a fallback threshold of 0.5 or an adaptive top-\(k\) selection is applied. The loss is
	\begin{equation}\label{eq:supp-peak}
		\mathcal{L}_{\mathrm{peak}}
		=
		-\frac{1}{|\Omega_{\mathrm{peak}}|}
		\sum_{p\in\Omega_{\mathrm{peak}}}
		\log\sigma(z_p).
	\end{equation}
	
	To suppress false-positive responses in difficult background regions, the hard background set \(\Omega_{\mathrm{hard}}\) is selected as the top-ranked predicted logits among pixels satisfying \(y_p<0.05\). The hard background suppression loss is
	\begin{equation}\label{eq:supp-hard-bg}
		\mathcal{L}_{\mathrm{hardBG}}
		=
		-\frac{1}{|\Omega_{\mathrm{hard}}|}
		\sum_{p\in\Omega_{\mathrm{hard}}}
		\log(1-\sigma(z_p)).
	\end{equation}
	
	The ranking loss encourages lesion-center responses to exceed hard background activations. The peak anchor score is
	\[
	a_{\mathrm{peak}}
	=
	\frac{
		\sum_{p\in\Omega_{\mathrm{peak}}}z_p y_p
	}{
		\sum_{p\in\Omega_{\mathrm{peak}}}y_p+\epsilon
	}.
	\]
	Then,
	\begin{equation}\label{eq:supp-rank}
		\mathcal{L}_{\mathrm{rank}}
		=
		\frac{1}{|\Omega_{\mathrm{hard}}|}
		\sum_{q\in\Omega_{\mathrm{hard}}}
		\operatorname{softplus}
		(z_q-a_{\mathrm{peak}}+m),
	\end{equation}
	where \(m=2.0\) is the ranking margin.
	
	The mass calibration loss constrains the total predicted heatmap response:
	\begin{equation}\label{eq:supp-mass}
		\mathcal{L}_{\mathrm{mass}}
		=
		\operatorname{SmoothL1}
		\left(
		\log(\sum_p\sigma(z_p)),
		\log(\sum_p y_p)
		\right).
	\end{equation}
	
	The hard SFD gate uses the following separately defined per-sample detection risk and activates only when the student risk exceeds the teacher risk by more than 0.02:
	\begin{equation}\label{eq:sfd-det-risk}
		\begin{aligned}
			R_{\mathrm{det},i}
			&=
			\mathcal{L}_{\mathrm{BCE},i}
			+0.03\mathcal{L}_{\mathrm{KL},i}
			+0.05\mathcal{L}_{\mathrm{MSE},i}
			+d_{\mathrm{peak},i},\\
			g_i
			&=\mathbf{1}\!\left\{R_{\mathrm{det},i}^{S}-R_{\mathrm{det},i}^{T}>0.02\right\}.
		\end{aligned}
	\end{equation}
	Here, \(\mathcal{L}_{\mathrm{BCE},i}\) denotes the standard per-sample binary cross-entropy used only for risk estimation in the SFD gate. It is intentionally different from the balanced BCE objective \(\mathcal{L}_{\mathrm{balBCE}}\) used for detector optimization in Eq.~\eqref{eq:sfd-det}. The term \(d_{\mathrm{peak},i}\) is the Euclidean distance between the highest predicted and ground-truth heatmap peaks, normalized by the image diagonal. For SFD training, the distillation coefficient is initialized as zero during the first 20 warm-up epochs and linearly increased over the subsequent 30 epochs until reaching the final value. This warm-up and ramp strategy is applied to both BraTS segmentation and DeepLesion centre-point detection experiments.
	
	For DeepLesion, the spatially weighted and channel-normalized feature-distillation loss at layer \(\ell\) is
	\begin{equation}\label{eq:sfd-det-distill}
		\mathcal{L}_{\mathrm{det\text{-}distill}}^{\ell}
		=
		\frac{
			\sum_i g_i
			\left\langle
			W_i^{\ell},
			\left\|
			\operatorname{norm}_{c}(F_{S,i}^{\ell})
			-
			\operatorname{norm}_{c}
			\left(\mathrm{sg}[F_{T,i}^{\ell}]\right)
			\right\|_2^2
			\right\rangle
		}{
			\sum_i g_i
			\sum_{h,w}W_{i,h,w}^{\ell}C_\ell+\epsilon
		}.
	\end{equation}
	Here, \(W_i^\ell\) denotes the spatial weighting map for sample \(i\) at layer \(\ell\), \(\operatorname{norm}_c\) denotes normalization along the channel dimension, \(\mathrm{sg}\) is the stop-gradient operator, \(\langle\cdot,\cdot\rangle\) denotes summation of the element-wise product over spatial locations, \(C_\ell\) is the number of channels, and \(\epsilon\) is a numerical-stability constant. If no sample in a batch activates the gate, the numerator is zero and the distillation term evaluates to zero. Layer-wise terms are combined over the selected feature layers as specified by the main-text SFD objective.
	
	\subsection{DeepLesion detection evaluation protocol}
	For DeepLesion evaluation, FROC sensitivity and conditional mean matched localization error were calculated using separate candidate-selection protocols. For FROC sensitivity, predicted peaks were extracted using 9$\times$9 non-maximum suppression (NMS), with an absolute score threshold of \(10^{-4}\) and a maximum of 50 candidates per image. Predictions were matched to ground-truth lesion centres within a 15-pixel radius. Patient-level sensitivity was evaluated at false-positive budgets of 0.5, 1, and 2 per image.
	
	For conditional localization error, the same 9$\times$9 NMS was applied, but at most three predicted peaks per image were retained, using a relative threshold of 0.8 times the maximum predicted heatmap value. Ground-truth peaks were extracted using a threshold of 0.99, with at most three peaks per image. Predicted and ground-truth peaks were matched one-to-one in ascending order of Euclidean distance within a 15-pixel radius. The mean distance between successfully matched pairs was calculated for each image and then averaged across images with valid matches for each patient. Images without valid matches were excluded from this conditional metric. The paired localization-error analysis included only the 528 patients with valid matched detections for both methods. Both the baseline detector and SFD used identical evaluation protocols.
	
	\clearpage
	\begin{landscape}
		\pagestyle{empty}
		\thispagestyle{empty}
		\AddToShipoutPictureFG{%
			\AtPageLowerLeft{%
				\put(\LenToUnit{\dimexpr\paperwidth-20pt\relax},%
				\LenToUnit{0.5\paperheight}){%
					\rotatebox{90}{\makebox[0pt][c]{\normalfont\normalsize\thepage}}%
				}%
			}%
		}
		\section{Supplementary Statistical Analyses}
		\label{supp:statistical-analyses}
		
		All analyses use paired observations at the independent case or patient level defined for the corresponding experiment. The tables report paired effect estimates together with pointwise 95\% confidence intervals (CIs). For Supplementary Table~\ref{tab:afd_sig_appendix}, mean paired differences are reported to align with the cohort-mean bars, and median paired differences with percentile-bootstrap CIs are shown separately. FID is a dataset-level metric and is therefore reported descriptively in the main text rather than included in the case-level paired analyses. Holm correction is applied separately within each table, with 36, 12, 8, 36, 8 and 5 comparisons in Tables~\ref{tab:stat_fig2a}--\ref{tab:sfd_det_sig_appendix}, respectively. Each table constitutes one family, including all listed metrics, reference-method comparisons, and translation directions where applicable; family identifiers and test counts are specified in the captions. Statistical significance is defined by Holm-adjusted \(p<0.05\) within the corresponding family. All CIs remain pointwise and unadjusted for multiplicity. Significant adjusted results are reported using threshold ranges, and nonsignificant adjusted results are reported numerically. This procedure controls family-wise error within each table, not across the combined set of all six tables.
		
		\begingroup
		\setlength{\LTleft}{\fill}
		\setlength{\LTright}{\fill}
		\setlength{\tabcolsep}{2pt}
		\renewcommand{\arraystretch}{0.96}
		\captionsetup[table]{font=footnotesize}
		
		\begin{longtable}{
				S{1.65cm}
				L{5.10cm}
				S{1.10cm}
				S{0.65cm}
				S{1.50cm}
				S{3.20cm}
				S{1.25cm}
				S{1.90cm}}
			\caption{Case-level paired analyses for the held-out 219-case official BraTS 2021 validation cohort used exclusively for final evaluation shown in Fig.~\ref{fig2}a. Each case contributed one value per metric. Paired \(t\)-tests were used. The paired mean difference was defined as \(\Delta=\mathrm{FDDT\ variant}-\mathrm{reference}\); positive values favor FDDT for SSIM and PSNR, whereas negative values favor FDDT for LPIPS. The 95\% CIs are pointwise CIs for the mean paired differences. All comparisons belong to family S1 (\(m=36\)).}
			\label{tab:stat_fig2a}\\
			
			\toprule
			Task & Comparison & Metric & \(n\) & Mean \(\Delta\) & 95\% CI of \(\Delta\) & Raw \(p\) & \shortstack{Holm-adjusted\\\(p\)}\\
			\midrule
			\endfirsthead
			
			\noalign{\vskip 8pt}
			\multicolumn{8}{c}{\textit{Supplementary Table \thetable\ continued}}\\
			\toprule
			Task & Comparison & Metric & \(n\) & Mean \(\Delta\) & 95\% CI of \(\Delta\) & Raw \(p\) & \shortstack{Holm-adjusted\\\(p\)}\\
			\midrule
			\endhead
			
			\midrule
			\multicolumn{8}{r}{\textit{Continued on next page}}\\
			\endfoot
			
			\bottomrule
			\endlastfoot
			
			T1\(\rightarrow\)T1ce & CycleGAN+FDDT vs.\ CycleGAN & SSIM & 219 & \(0.02088\) & \([0.01812, 0.02363]\) & \(p<0.001\) & \(p<0.001\)\\
			T1\(\rightarrow\)T1ce & CycleGAN+FDDT vs.\ CycleGAN & PSNR & 219 & \(1.40474\) & \([0.91617, 1.89331]\) & \(p<0.001\) & \(p<0.001\)\\
			T1\(\rightarrow\)T1ce & CycleGAN+FDDT vs.\ CycleGAN & LPIPS & 219 & \(-0.01665\) & \([-0.01909, -0.01421]\) & \(p<0.001\) & \(p<0.001\)\\
			T1\(\rightarrow\)T1ce & CycleGAN+FDDT vs.\ CycleGAN+FDIT & SSIM & 219 & \(0.03591\) & \([0.03262, 0.03921]\) & \(p<0.001\) & \(p<0.001\)\\
			T1\(\rightarrow\)T1ce & CycleGAN+FDDT vs.\ CycleGAN+FDIT & PSNR & 219 & \(2.43778\) & \([1.88086, 2.99470]\) & \(p<0.001\) & \(p<0.001\)\\
			T1\(\rightarrow\)T1ce & CycleGAN+FDDT vs.\ CycleGAN+FDIT & LPIPS & 219 & \(-0.02240\) & \([-0.02506, -0.01973]\) & \(p<0.001\) & \(p<0.001\)\\
			T1\(\rightarrow\)T1ce & CUT+FDDT vs.\ CUT & SSIM & 219 & \(0.01747\) & \([0.01595, 0.01898]\) & \(p<0.001\) & \(p<0.001\)\\
			T1\(\rightarrow\)T1ce & CUT+FDDT vs.\ CUT & PSNR & 219 & \(2.44851\) & \([2.13579, 2.76123]\) & \(p<0.001\) & \(p<0.001\)\\
			T1\(\rightarrow\)T1ce & CUT+FDDT vs.\ CUT & LPIPS & 219 & \(-0.00309\) & \([-0.00424, -0.00194]\) & \(p<0.001\) & \(p<0.001\)\\
			T1\(\rightarrow\)T1ce & CUT+FDDT vs.\ CUT+FDIT & SSIM & 219 & \(0.01368\) & \([0.00987, 0.01749]\) & \(p<0.001\) & \(p<0.001\)\\
			T1\(\rightarrow\)T1ce & CUT+FDDT vs.\ CUT+FDIT & PSNR & 219 & \(3.27930\) & \([2.74903, 3.80957]\) & \(p<0.001\) & \(p<0.001\)\\
			T1\(\rightarrow\)T1ce & CUT+FDDT vs.\ CUT+FDIT & LPIPS & 219 & \(-0.01263\) & \([-0.01442, -0.01084]\) & \(p<0.001\) & \(p<0.001\)\\
			T1\(\rightarrow\)T1ce & SynDiff+FDDT vs.\ SynDiff & SSIM & 219 & \(0.01667\) & \([0.01365, 0.01968]\) & \(p<0.001\) & \(p<0.001\)\\
			T1\(\rightarrow\)T1ce & SynDiff+FDDT vs.\ SynDiff & PSNR & 219 & \(3.03724\) & \([2.50765, 3.56682]\) & \(p<0.001\) & \(p<0.001\)\\
			T1\(\rightarrow\)T1ce & SynDiff+FDDT vs.\ SynDiff & LPIPS & 219 & \(-0.00648\) & \([-0.00835, -0.00461]\) & \(p<0.001\) & \(p<0.001\)\\
			T1\(\rightarrow\)T1ce & SynDiff+FDDT vs.\ SynDiff+FDIT & SSIM & 219 & \(0.02910\) & \([0.02647, 0.03173]\) & \(p<0.001\) & \(p<0.001\)\\
			T1\(\rightarrow\)T1ce & SynDiff+FDDT vs.\ SynDiff+FDIT & PSNR & 219 & \(2.33024\) & \([1.85220, 2.80828]\) & \(p<0.001\) & \(p<0.001\)\\
			T1\(\rightarrow\)T1ce & SynDiff+FDDT vs.\ SynDiff+FDIT & LPIPS & 219 & \(-0.00600\) & \([-0.00802, -0.00398]\) & \(p<0.001\) & \(p<0.001\)\\
			\addlinespace[2pt]
			T2\(\rightarrow\)FLAIR & CycleGAN+FDDT vs.\ CycleGAN & SSIM & 219 & \(0.02542\) & \([0.02356, 0.02728]\) & \(p<0.001\) & \(p<0.001\)\\
			T2\(\rightarrow\)FLAIR & CycleGAN+FDDT vs.\ CycleGAN & PSNR & 219 & \(1.42142\) & \([1.28295, 1.55989]\) & \(p<0.001\) & \(p<0.001\)\\
			T2\(\rightarrow\)FLAIR & CycleGAN+FDDT vs.\ CycleGAN & LPIPS & 219 & \(-0.01598\) & \([-0.01823, -0.01373]\) & \(p<0.001\) & \(p<0.001\)\\
			T2\(\rightarrow\)FLAIR & CycleGAN+FDDT vs.\ CycleGAN+FDIT & SSIM & 219 & \(0.02485\) & \([0.02337, 0.02633]\) & \(p<0.001\) & \(p<0.001\)\\
			T2\(\rightarrow\)FLAIR & CycleGAN+FDDT vs.\ CycleGAN+FDIT & PSNR & 219 & \(1.28707\) & \([1.14258, 1.43155]\) & \(p<0.001\) & \(p<0.001\)\\
			T2\(\rightarrow\)FLAIR & CycleGAN+FDDT vs.\ CycleGAN+FDIT & LPIPS & 219 & \(-0.01412\) & \([-0.01627, -0.01196]\) & \(p<0.001\) & \(p<0.001\)\\
			T2\(\rightarrow\)FLAIR & CUT+FDDT vs.\ CUT & SSIM & 219 & \(0.03198\) & \([0.03039, 0.03358]\) & \(p<0.001\) & \(p<0.001\)\\
			T2\(\rightarrow\)FLAIR & CUT+FDDT vs.\ CUT & PSNR & 219 & \(1.46943\) & \([1.19353, 1.74533]\) & \(p<0.001\) & \(p<0.001\)\\
			T2\(\rightarrow\)FLAIR & CUT+FDDT vs.\ CUT & LPIPS & 219 & \(-0.00559\) & \([-0.00710, -0.00409]\) & \(p<0.001\) & \(p<0.001\)\\
			T2\(\rightarrow\)FLAIR & CUT+FDDT vs.\ CUT+FDIT & SSIM & 219 & \(0.01894\) & \([0.01766, 0.02022]\) & \(p<0.001\) & \(p<0.001\)\\
			T2\(\rightarrow\)FLAIR & CUT+FDDT vs.\ CUT+FDIT & PSNR & 219 & \(2.12608\) & \([1.92741, 2.32475]\) & \(p<0.001\) & \(p<0.001\)\\
			T2\(\rightarrow\)FLAIR & CUT+FDDT vs.\ CUT+FDIT & LPIPS & 219 & \(-0.01148\) & \([-0.01268, -0.01028]\) & \(p<0.001\) & \(p<0.001\)\\
			T2\(\rightarrow\)FLAIR & SynDiff+FDDT vs.\ SynDiff & SSIM & 219 & \(0.02795\) & \([0.02322, 0.03268]\) & \(p<0.001\) & \(p<0.001\)\\
			T2\(\rightarrow\)FLAIR & SynDiff+FDDT vs.\ SynDiff & PSNR & 219 & \(1.49087\) & \([1.20410, 1.77764]\) & \(p<0.001\) & \(p<0.001\)\\
			T2\(\rightarrow\)FLAIR & SynDiff+FDDT vs.\ SynDiff & LPIPS & 219 & \(-0.01248\) & \([-0.01421, -0.01074]\) & \(p<0.001\) & \(p<0.001\)\\
			T2\(\rightarrow\)FLAIR & SynDiff+FDDT vs.\ SynDiff+FDIT & SSIM & 219 & \(0.01888\) & \([0.01665, 0.02112]\) & \(p<0.001\) & \(p<0.001\)\\
			T2\(\rightarrow\)FLAIR & SynDiff+FDDT vs.\ SynDiff+FDIT & PSNR & 219 & \(0.92301\) & \([0.68860, 1.15742]\) & \(p<0.001\) & \(p<0.001\)\\
			T2\(\rightarrow\)FLAIR & SynDiff+FDDT vs.\ SynDiff+FDIT & LPIPS & 219 & \(-0.01716\) & \([-0.01907, -0.01525]\) & \(p<0.001\) & \(p<0.001\)\\
		\end{longtable}

		\clearpage

		\begin{longtable}{
				S{1.50cm}
				L{5.25cm}
				S{1.10cm}
				S{0.65cm}
				S{1.50cm}
				S{3.20cm}
				S{1.25cm}
				S{1.90cm}}
			\caption{Case-level paired analyses for the 103-case held-out SynthRAD2025 CT\(\rightarrow\)MR test partition shown in Fig.~\ref{fig2}b. Each case contributed one value per metric. Paired \(t\)-tests were used. The paired mean difference was defined as \(\Delta=\mathrm{FDDT\ variant}-\mathrm{reference}\); positive values favor FDDT. All comparisons belong to family S2 (\(m=12\)).}
			\label{tab:stat_fig2b}\\
			
			\toprule
			Task & Comparison & Metric & \(n\) & Mean \(\Delta\) & 95\% CI of \(\Delta\) & Raw \(p\) & \shortstack{Holm-adjusted\\\(p\)}\\
			\midrule
			\endfirsthead
			
			\noalign{\vskip 8pt}
			\multicolumn{8}{c}{\textit{Supplementary Table \thetable\ continued}}\\
			\toprule
			Task & Comparison & Metric & \(n\) & Mean \(\Delta\) & 95\% CI of \(\Delta\) & Raw \(p\) & \shortstack{Holm-adjusted\\\(p\)}\\
			\midrule
			\endhead
			
			\midrule
			\multicolumn{8}{r}{\textit{Continued on next page}}\\
			\endfoot
			
			\bottomrule
			\endlastfoot
			
			CT\(\rightarrow\)MR & CycleGAN+FDDT vs.\ CycleGAN & SSIM & 103 & \(0.01745\) & \([0.01522, 0.01968]\) & \(p<0.001\) & \(p<0.001\)\\
			CT\(\rightarrow\)MR & CycleGAN+FDDT vs.\ CycleGAN & PSNR & 103 & \(0.42025\) & \([0.30006, 0.54044]\) & \(p<0.001\) & \(p<0.001\)\\
			CT\(\rightarrow\)MR & CycleGAN+FDDT vs.\ CycleGAN+FDIT & SSIM & 103 & \(0.02103\) & \([0.01768, 0.02438]\) & \(p<0.001\) & \(p<0.001\)\\
			CT\(\rightarrow\)MR & CycleGAN+FDDT vs.\ CycleGAN+FDIT & PSNR & 103 & \(1.32802\) & \([1.17242, 1.48361]\) & \(p<0.001\) & \(p<0.001\)\\
			CT\(\rightarrow\)MR & CUT+FDDT vs.\ CUT & SSIM & 103 & \(0.01489\) & \([0.01218, 0.01760]\) & \(p<0.001\) & \(p<0.001\)\\
			CT\(\rightarrow\)MR & CUT+FDDT vs.\ CUT & PSNR & 103 & \(0.45838\) & \([0.22098, 0.69579]\) & \(p<0.001\) & \(p<0.001\)\\
			CT\(\rightarrow\)MR & CUT+FDDT vs.\ CUT+FDIT & SSIM & 103 & \(0.03303\) & \([0.02858, 0.03748]\) & \(p<0.001\) & \(p<0.001\)\\
			CT\(\rightarrow\)MR & CUT+FDDT vs.\ CUT+FDIT & PSNR & 103 & \(1.60556\) & \([1.31523, 1.89590]\) & \(p<0.001\) & \(p<0.001\)\\
			CT\(\rightarrow\)MR & SynDiff+FDDT vs.\ SynDiff & SSIM & 103 & \(0.01220\) & \([0.01003, 0.01436]\) & \(p<0.001\) & \(p<0.001\)\\
			CT\(\rightarrow\)MR & SynDiff+FDDT vs.\ SynDiff & PSNR & 103 & \(0.56523\) & \([0.42424, 0.70622]\) & \(p<0.001\) & \(p<0.001\)\\
			CT\(\rightarrow\)MR & SynDiff+FDDT vs.\ SynDiff+FDIT & SSIM & 103 & \(0.01424\) & \([0.01213, 0.01635]\) & \(p<0.001\) & \(p<0.001\)\\
			CT\(\rightarrow\)MR & SynDiff+FDDT vs.\ SynDiff+FDIT & PSNR & 103 & \(1.09554\) & \([0.91646, 1.27462]\) & \(p<0.001\) & \(p<0.001\)\\
		\end{longtable}

		\clearpage

		\begin{longtable}{
				S{1.30cm}
				L{4.35cm}
				S{1.05cm}
				S{0.60cm}
				S{1.45cm}
				S{3.50cm}
				S{1.25cm}
				S{1.90cm}
			}
			\caption{Case-level paired analyses for the IXI AFD ablation experiments shown in Fig.~\ref{fig3}b. Each of the 88 cases contributed one value per metric. The mean paired difference, \(\Delta=\mathrm{AFD\text{-}enabled\ variant}-\mathrm{reference}\), is reported to align the effect estimate with the cohort-mean bars. Paired Wilcoxon signed-rank tests were used; the median paired difference and its 95\% percentile-bootstrap CI, calculated using 5,000 case-level paired resamples with replacement (seed 42), are reported separately as the nonparametric effect summary. Raw \(p\)-values are from the Wilcoxon tests; the final column reports the corresponding Holm-adjusted \(p\)-values. Positive values favor the AFD-enabled variant. All comparisons belong to family S3 (\(m=8\)).}
			\label{tab:afd_sig_appendix}\\
			
			\toprule
			Task & Comparison & Metric & \(n\) & Mean \(\Delta\) & \shortstack{Median \(\Delta\)\\(95\% CI)} & Raw \(p\) & \shortstack{Holm-adjusted\\\(p\)}\\
			\midrule
			\endfirsthead
			
			\noalign{\vskip 8pt}
			\multicolumn{8}{c}{\textit{Supplementary Table \thetable\ continued}}\\
			\toprule
			Task & Comparison & Metric & \(n\) & Mean \(\Delta\) & \shortstack{Median \(\Delta\)\\(95\% CI)} & Raw \(p\) & \shortstack{Holm-adjusted\\\(p\)}\\
			\midrule
			\endhead
			
			\midrule
			\multicolumn{8}{r}{\textit{Continued on next page}}\\
			\endfoot
			
			\bottomrule
			\endlastfoot
			
			PD\(\rightarrow\)T2 & CycleGAN+FDDT+AFD vs.\ CycleGAN & SSIM & 88 & \(0.025449\) & \shortstack{\(0.028984\)\\\([0.012512, 0.042527]\)} & \(p<0.001\) & \(p<0.01\)\\
			PD\(\rightarrow\)T2 & CycleGAN+FDDT+AFD vs.\ CycleGAN & PSNR & 88 & \(1.287684\) & \shortstack{\(1.331045\)\\\([0.425650, 1.872452]\)} & \(p<0.001\) & \(p<0.01\)\\
			PD\(\rightarrow\)T2 & CycleGAN+FDDT+AFD vs.\ CycleGAN+FDDT & SSIM & 88 & \(0.013659\) & \shortstack{\(0.018983\)\\\([0.015480, 0.022335]\)} & \(p<0.001\) & \(p<0.01\)\\
			PD\(\rightarrow\)T2 & CycleGAN+FDDT+AFD vs.\ CycleGAN+FDDT & PSNR & 88 & \(0.718979\) & \shortstack{\(0.552907\)\\\([0.354597, 1.004461]\)} & \(p<0.001\) & \(p<0.01\)\\
			\addlinespace[2pt]
			T2\(\rightarrow\)PD & CycleGAN+FDDT+AFD vs.\ CycleGAN & SSIM & 88 & \(0.007752\) & \shortstack{\(0.006950\)\\\([0.004337, 0.009257]\)} & \(p<0.001\) & \(p<0.01\)\\
			T2\(\rightarrow\)PD & CycleGAN+FDDT+AFD vs.\ CycleGAN & PSNR & 88 & \(0.281004\) & \shortstack{\(0.299682\)\\\([0.161102, 0.340695]\)} & \(p<0.001\) & \(p<0.01\)\\
			T2\(\rightarrow\)PD & CycleGAN+FDDT+AFD vs.\ CycleGAN+FDDT & SSIM & 88 & \(0.011595\) & \shortstack{\(0.015863\)\\\([0.009805, 0.024749]\)} & \(p<0.001\) & \(p<0.01\)\\
			T2\(\rightarrow\)PD & CycleGAN+FDDT+AFD vs.\ CycleGAN+FDDT & PSNR & 88 & \(1.114905\) & \shortstack{\(0.684201\)\\\([0.507771, 1.213841]\)} & \(p<0.001\) & \(p<0.01\)\\
		\end{longtable}

		\clearpage

		\begin{longtable}{
				S{1.45cm}
				L{4.90cm}
				S{0.90cm}
				S{0.60cm}
				S{1.20cm}
				S{2.80cm}
				S{1.25cm}
				S{1.90cm}
			}
			\caption{Patient-level paired analyses of SSIM, PSNR, and LPIPS for external validation of the BraTS 2021-trained translators, evaluated without retraining in 87 BraTS 2024 patients comprising 187 longitudinal MRI examinations (Fig.~\ref{fig5}). Examination-level metrics were averaged within patient across available time points. Paired \(t\)-tests were used. The paired mean difference was defined as \(\Delta=\mathrm{FDDT\ variant}-\mathrm{reference}\); positive values favor FDDT for SSIM and PSNR, whereas negative values favor FDDT for LPIPS. The 95\% CIs are pointwise CIs for the mean paired differences. All comparisons belong to family S4 (\(m=36\)). All 36 mean differences favour FDDT; 35 pointwise CIs exclude zero and 30 comparisons remain significant after Holm correction.}
			\label{tab:brats2024_sig_appendix}\\
			
			\toprule
			Task & Comparison & Metric & \(n\) & Mean \(\Delta\) & 95\% CI of \(\Delta\) & Raw \(p\) & \shortstack{Holm-adjusted\\\(p\)}\\
			\midrule
			\endfirsthead
			
			\noalign{\vskip 8pt}
			\multicolumn{8}{c}{\textit{Supplementary Table \thetable\ continued}}\\
			\toprule
			Task & Comparison & Metric & \(n\) & Mean \(\Delta\) & 95\% CI of \(\Delta\) & Raw \(p\) & \shortstack{Holm-adjusted\\\(p\)}\\
			\midrule
			\endhead
			
			\midrule
			\multicolumn{8}{r}{\textit{Continued on next page}}\\
			\endfoot
			
			\bottomrule
			\endlastfoot
			
			T1\(\rightarrow\)T1ce & CycleGAN+FDDT vs.\ CycleGAN & SSIM & 87 & \(0.00791\) & \([0.00633, 0.00949]\) & \(p<0.001\) & \(p<0.001\)\\
			T1\(\rightarrow\)T1ce & CycleGAN+FDDT vs.\ CycleGAN & PSNR & 87 & \(0.31588\) & \([0.15110, 0.48066]\) & \(p<0.001\) & \(p<0.01\)\\
			T1\(\rightarrow\)T1ce & CycleGAN+FDDT vs.\ CycleGAN & LPIPS & 87 & \(-0.01019\) & \([-0.01259, -0.00779]\) & \(p<0.001\) & \(p<0.001\)\\
			T1\(\rightarrow\)T1ce & CycleGAN+FDDT vs.\ CycleGAN+FDIT & SSIM & 87 & \(0.00679\) & \([0.00431, 0.00927]\) & \(p<0.001\) & \(p<0.001\)\\
			T1\(\rightarrow\)T1ce & CycleGAN+FDDT vs.\ CycleGAN+FDIT & PSNR & 87 & \(0.43504\) & \([0.19955, 0.67053]\) & \(p<0.001\) & \(p<0.01\)\\
			T1\(\rightarrow\)T1ce & CycleGAN+FDDT vs.\ CycleGAN+FDIT & LPIPS & 87 & \(-0.00588\) & \([-0.00774, -0.00402]\) & \(p<0.001\) & \(p<0.001\)\\
			T1\(\rightarrow\)T1ce & CUT+FDDT vs.\ CUT & SSIM & 87 & \(0.01014\) & \([0.00908, 0.01120]\) & \(p<0.001\) & \(p<0.001\)\\
			T1\(\rightarrow\)T1ce & CUT+FDDT vs.\ CUT & PSNR & 87 & \(0.54623\) & \([0.36389, 0.72857]\) & \(p<0.001\) & \(p<0.001\)\\
			T1\(\rightarrow\)T1ce & CUT+FDDT vs.\ CUT & LPIPS & 87 & \(-0.00923\) & \([-0.01030, -0.00816]\) & \(p<0.001\) & \(p<0.001\)\\
			T1\(\rightarrow\)T1ce & CUT+FDDT vs.\ CUT+FDIT & SSIM & 87 & \(0.00749\) & \([0.00494, 0.01004]\) & \(p<0.001\) & \(p<0.001\)\\
			T1\(\rightarrow\)T1ce & CUT+FDDT vs.\ CUT+FDIT & PSNR & 87 & \(0.52794\) & \([0.29578, 0.76010]\) & \(p<0.001\) & \(p<0.001\)\\
			T1\(\rightarrow\)T1ce & CUT+FDDT vs.\ CUT+FDIT & LPIPS & 87 & \(-0.00564\) & \([-0.00743, -0.00385]\) & \(p<0.001\) & \(p<0.001\)\\
			T1\(\rightarrow\)T1ce & SynDiff+FDDT vs.\ SynDiff & SSIM & 87 & \(0.00649\) & \([0.00522, 0.00776]\) & \(p<0.001\) & \(p<0.001\)\\
			T1\(\rightarrow\)T1ce & SynDiff+FDDT vs.\ SynDiff & PSNR & 87 & \(0.44688\) & \([0.27271, 0.62105]\) & \(p<0.001\) & \(p<0.001\)\\
			T1\(\rightarrow\)T1ce & SynDiff+FDDT vs.\ SynDiff & LPIPS & 87 & \(-0.00980\) & \([-0.01113, -0.00847]\) & \(p<0.001\) & \(p<0.001\)\\
			T1\(\rightarrow\)T1ce & SynDiff+FDDT vs.\ SynDiff+FDIT & SSIM & 87 & \(0.00207\) & \([0.00054, 0.00360]\) & \(p<0.01\) & \(0.05156\)\\
			T1\(\rightarrow\)T1ce & SynDiff+FDDT vs.\ SynDiff+FDIT & PSNR & 87 & \(0.32935\) & \([0.00772, 0.65098]\) & \(p<0.05\) & \(0.15595\)\\
			T1\(\rightarrow\)T1ce & SynDiff+FDDT vs.\ SynDiff+FDIT & LPIPS & 87 & \(-0.00563\) & \([-0.00686, -0.00440]\) & \(p<0.001\) & \(p<0.001\)\\
			\addlinespace[2pt]
			T2\(\rightarrow\)FLAIR & CycleGAN+FDDT vs.\ CycleGAN & SSIM & 87 & \(0.00689\) & \([0.00469, 0.00909]\) & \(p<0.001\) & \(p<0.001\)\\
			T2\(\rightarrow\)FLAIR & CycleGAN+FDDT vs.\ CycleGAN & PSNR & 87 & \(0.98179\) & \([0.58286, 1.38072]\) & \(p<0.001\) & \(p<0.001\)\\
			T2\(\rightarrow\)FLAIR & CycleGAN+FDDT vs.\ CycleGAN & LPIPS & 87 & \(-0.00548\) & \([-0.01073, -0.00023]\) & \(p<0.05\) & \(0.15595\)\\
			T2\(\rightarrow\)FLAIR & CycleGAN+FDDT vs.\ CycleGAN+FDIT & SSIM & 87 & \(0.00406\) & \([0.00021, 0.00791]\) & \(p<0.05\) & \(0.15595\)\\
			T2\(\rightarrow\)FLAIR & CycleGAN+FDDT vs.\ CycleGAN+FDIT & PSNR & 87 & \(1.06825\) & \([0.53938, 1.59712]\) & \(p<0.001\) & \(p<0.01\)\\
			T2\(\rightarrow\)FLAIR & CycleGAN+FDDT vs.\ CycleGAN+FDIT & LPIPS & 87 & \(-0.00605\) & \([-0.00958, -0.00252]\) & \(p<0.01\) & \(p<0.01\)\\
			T2\(\rightarrow\)FLAIR & CUT+FDDT vs.\ CUT & SSIM & 87 & \(0.00454\) & \([0.00053, 0.00855]\) & \(p<0.05\) & \(0.13478\)\\
			T2\(\rightarrow\)FLAIR & CUT+FDDT vs.\ CUT & PSNR & 87 & \(0.58002\) & \([0.42361, 0.73643]\) & \(p<0.001\) & \(p<0.001\)\\
			T2\(\rightarrow\)FLAIR & CUT+FDDT vs.\ CUT & LPIPS & 87 & \(-0.00772\) & \([-0.01175, -0.00369]\) & \(p<0.001\) & \(p<0.01\)\\
			T2\(\rightarrow\)FLAIR & CUT+FDDT vs.\ CUT+FDIT & SSIM & 87 & \(0.00888\) & \([0.00644, 0.01132]\) & \(p<0.001\) & \(p<0.001\)\\
			T2\(\rightarrow\)FLAIR & CUT+FDDT vs.\ CUT+FDIT & PSNR & 87 & \(0.61530\) & \([0.40566, 0.82494]\) & \(p<0.001\) & \(p<0.001\)\\
			T2\(\rightarrow\)FLAIR & CUT+FDDT vs.\ CUT+FDIT & LPIPS & 87 & \(-0.00429\) & \([-0.00930, 0.00072]\) & \(0.09232\) & \(0.15595\)\\
			T2\(\rightarrow\)FLAIR & SynDiff+FDDT vs.\ SynDiff & SSIM & 87 & \(0.00840\) & \([0.00720, 0.00960]\) & \(p<0.001\) & \(p<0.001\)\\
			T2\(\rightarrow\)FLAIR & SynDiff+FDDT vs.\ SynDiff & PSNR & 87 & \(0.39745\) & \([0.27141, 0.52349]\) & \(p<0.001\) & \(p<0.001\)\\
			T2\(\rightarrow\)FLAIR & SynDiff+FDDT vs.\ SynDiff & LPIPS & 87 & \(-0.00559\) & \([-0.00806, -0.00312]\) & \(p<0.001\) & \(p<0.001\)\\
			T2\(\rightarrow\)FLAIR & SynDiff+FDDT vs.\ SynDiff+FDIT & SSIM & 87 & \(0.02441\) & \([0.01594, 0.03288]\) & \(p<0.001\) & \(p<0.001\)\\
			T2\(\rightarrow\)FLAIR & SynDiff+FDDT vs.\ SynDiff+FDIT & PSNR & 87 & \(0.44382\) & \([0.23175, 0.65589]\) & \(p<0.001\) & \(p<0.001\)\\
			T2\(\rightarrow\)FLAIR & SynDiff+FDDT vs.\ SynDiff+FDIT & LPIPS & 87 & \(-0.00630\) & \([-0.01006, -0.00254]\) & \(p<0.01\) & \(p<0.01\)\\
		\end{longtable}

		\clearpage

		\begin{longtable}{
				L{5.30cm}
				S{1.45cm}
				S{0.65cm}
				S{1.25cm}
				S{2.75cm}
				S{1.25cm}
				S{1.90cm}
			}
			\caption{Case-level paired analyses for the 249-case held-out BraTS 2021 tumour-containing-slice segmentation test cohort shown in Fig.~\ref{fig4}a. Each case contributed one value per metric. Paired \(t\)-tests were used. The paired mean difference was defined as \(\Delta=\mathrm{SFD}-\mathrm{reference}\); positive values favor SFD. The 95\% CIs are pointwise CIs for the mean paired differences. All comparisons belong to family S5 (\(m=8\)). This table concerns the main 249-case segmentation comparison, not the separately trained 16-case FD-versus-SFD ablation.}
			\label{tab:sfd_seg_sig_appendix}\\
			
			\toprule
			Comparison & Metric & \(n\) & Mean \(\Delta\) & 95\% CI of \(\Delta\) & Raw \(p\) & \shortstack{Holm-adjusted\\\(p\)}\\
			\midrule
			\endfirsthead
			
			\noalign{\vskip 8pt}
			\multicolumn{7}{c}{\textit{Supplementary Table \thetable\ continued}}\\
			\toprule
			Comparison & Metric & \(n\) & Mean \(\Delta\) & 95\% CI of \(\Delta\) & Raw \(p\) & \shortstack{Holm-adjusted\\\(p\)}\\
			\midrule
			\endhead
			
			\midrule
			\multicolumn{7}{r}{\textit{Continued on next page}}\\
			\endfoot
			
			\bottomrule
			\endlastfoot
			
			SFD vs.\ UNet-SE-CA & Precision & 249 & \(0.05722\) & \([0.04784, 0.06659]\) & \(p<0.001\) & \(p<0.001\)\\
			SFD vs.\ UNet-SE-CA & Recall & 249 & \(0.00841\) & \([0.00002, 0.01680]\) & \(p<0.05\) & \(p<0.05\)\\
			SFD vs.\ UNet-SE-CA & Dice & 249 & \(0.03315\) & \([0.02672, 0.03957]\) & \(p<0.001\) & \(p<0.001\)\\
			SFD vs.\ UNet-SE-CA & IoU & 249 & \(0.04168\) & \([0.03485, 0.04850]\) & \(p<0.001\) & \(p<0.001\)\\
			SFD vs.\ Zhang et al. & Precision & 249 & \(0.02356\) & \([0.01536, 0.03177]\) & \(p<0.001\) & \(p<0.001\)\\
			SFD vs.\ Zhang et al. & Recall & 249 & \(0.02683\) & \([0.01955, 0.03411]\) & \(p<0.001\) & \(p<0.001\)\\
			SFD vs.\ Zhang et al. & Dice & 249 & \(0.02613\) & \([0.02052, 0.03174]\) & \(p<0.001\) & \(p<0.001\)\\
			SFD vs.\ Zhang et al. & IoU & 249 & \(0.03250\) & \([0.02661, 0.03840]\) & \(p<0.001\) & \(p<0.001\)\\
		\end{longtable}

		\clearpage

		\begin{longtable}{
				L{5.10cm}
				L{2.60cm}
				S{0.65cm}
				S{1.25cm}
				S{2.75cm}
				S{1.25cm}
				S{1.90cm}
			}
			\caption{Patient-level paired analyses for DeepLesion centre-point detection shown in Fig.~\ref{fig4}e. For each patient, sensitivity was calculated from confidence-ranked candidate peaks at false-positive budgets of 0.5, 1, and 2 FP per image, and the resulting patient-level values were used for paired analysis. Paired \(t\)-tests were used. The paired mean difference was defined as \(\Delta=\mathrm{CenterNet+SFD}-\mathrm{CenterNet}\). Positive values favor SFD for sensitivity, whereas negative values favor SFD for KL divergence and localization error. Sensitivity and KL-divergence analyses included all 663 test patients. Conditional mean localization error was evaluated in the 528 patients with valid matched detections for both methods; undetected lesions were excluded. The 95\% CIs are pointwise CIs for the mean paired differences. All comparisons belong to family S6 (\(m=5\)).}
			\label{tab:sfd_det_sig_appendix}\\
			
			\toprule
			Comparison & Metric & \(n\) & Mean \(\Delta\) & 95\% CI of \(\Delta\) & Raw \(p\) & \shortstack{Holm-adjusted\\\(p\)}\\
			\midrule
			\endfirsthead
			
			\noalign{\vskip 8pt}
			\multicolumn{7}{c}{\textit{Supplementary Table \thetable\ continued}}\\
			\toprule
			Comparison & Metric & \(n\) & Mean \(\Delta\) & 95\% CI of \(\Delta\) & Raw \(p\) & \shortstack{Holm-adjusted\\\(p\)}\\
			\midrule
			\endhead
			
			\midrule
			\multicolumn{7}{r}{\textit{Continued on next page}}\\
			\endfoot
			
			\bottomrule
			\endlastfoot
			
			CenterNet+SFD vs.\ CenterNet & Sensitivity at 0.5 FP/image & 663 & \(0.02817\) & \([0.00416, 0.05217]\) & \(p<0.05\) & \(p<0.05\)\\
			CenterNet+SFD vs.\ CenterNet & Sensitivity at 1 FP/image & 663 & \(0.02893\) & \([0.00474, 0.05312]\) & \(p<0.05\) & \(p<0.05\)\\
			CenterNet+SFD vs.\ CenterNet & Sensitivity at 2 FP/image & 663 & \(0.04142\) & \([0.01925, 0.06359]\) & \(p<0.001\) & \(p<0.01\)\\
			CenterNet+SFD vs.\ CenterNet & KL divergence & 663 & \(-0.43922\) & \([-0.55769, -0.32076]\) & \(p<0.001\) & \(p<0.001\)\\
			CenterNet+SFD vs.\ CenterNet & Conditional mean localization error (px) & 528 & \(-0.21628\) & \([-0.36810, -0.06446]\) & \(p<0.01\) & \(p<0.05\)\\
		\end{longtable}
		
		\endgroup
	\end{landscape}
	\ClearShipoutPictureFG
	\endgroup
	
\end{document}